\documentclass[nofootinbib,showpacs,preprintnumbers,amsmath,amssymb]{revtex4}
\usepackage{indentfirst}
\usepackage{hhline}
\usepackage{wrapfig, float}
\usepackage{array,longtable,tabularx}
\usepackage[T1]{fontenc}
\usepackage{multirow}
\usepackage{natbib}
\usepackage{graphicx}
\usepackage{dcolumn}
\usepackage{bm}
\usepackage{subfigure}
\usepackage[small]{caption2}
\usepackage{boxhandler}
\usepackage{color}
\usepackage{hyperref}
\usepackage{textcomp}
\renewcommand{\captionlabeldelim}{.}
\hypersetup{linkcolor=blue, citecolor=blue, urlcolor=blue, colorlinks=true}
\begin{document}



\title{Transport properties and Kohler's rule in $R$$_x$Lu$_{1-x}$B$_{12}$ solid solutions with $x$ $\leq$ 0.03: \\ do charge stripes really exist in metallic dodecaborides?}

\author{M.~Anisimov$^{1, 2}$}\email{anisimov.m.a@gmail.com}\author{N.~Samarin$^{1}$} \author{V. Krasnorussky$^{2}$}
\author{A.~Azarevich$^{1}$} \author{A.~Bogach$^{1}$} \author{V.~Glushkov$^{1}$}
\author{S.~Demishev$^{1, 2, 3}$} \author{V.~Voronov$^{1}$}\author{N.~Shitsevalova$^{4}$}\

\affiliation{$\phantom{x}^1$- Prokhorov General Physics Institute of the Russian Academy of Sciences, 38 Vavilov Street, 119991, Moscow, Russia}

\affiliation{$\phantom{x}^2$- Vereshchagin Institute for High Pressure Physics of RAS, 14 Kaluzhskoe Shosse, 142190 Troitsk, Russia}

 \affiliation{$\phantom{x}^3$- National Research University 'Higher School of Economics', 20 Myasnitskaya Street, 101000 Moscow, Russia}

\affiliation{$\phantom{x}^4$- Frantsevich Institute for Problems of Materials Science, National Academy of Sciences of Ukraine, 3 Krzhyzhanovsky Street, 03680, Kyiv, Ukraine}

\date{\today}
\begin{abstract}
Nonmagnetic metal LuB$_{12}$ is known to exhibit considerable transport anisotropy, which was explained in literature by different mechanisms including possible formation of dynamic charge stripes below the point $\sim$ 150~K. Here we study transport properties of solid solutions based on LuB$_{12}$ host compound with general formula $R$$_x$Lu$_{1-x}$B$_{12}$ ($R$$-$Dy, Er, Tm, Yb, Lu) and with $x$ $\leq$ 0.03. The experiment has been performed on single crystals of high quality in the temperature range 1.8 $-$ 300~K in magnetic fields up to 82~kOe. The application of several models to the analysis of zero-field resistivity is discussed. A phenomenological description of large positive quadratic component of transverse magnetoresistance $\Delta$$\rho$/$\rho$($H$) = $\mu_\textmd{D}^2$$H^2$, which dominates for all $R$$_x$Lu$_{1-x}$B$_{12}$ compounds under investigation, allows to estimate drift mobility exponential changes $\mu_\textmd{D}$ $\sim$ $T^{-\alpha}$ with the index $\alpha$ $\approx$ 0.95 $-$ 1.46. In order to check the existence of additional channel of scattering, caused by probable presence of dynamic charge stripes, we performed the study of the anisotropy of magnetoresistance in Dy$_{0.01}$Lu$_{0.99}$B$_{12}$ and Tm$_{0.03}$Lu$_{0.97}$B$_{12}$ compositions including the measurements of the field scans with different current and field geometries. The data obtained allow us to confirm the fulfillment of semi-empirical Kohler's rule in a wide interval of temperatures 30 $-$ 240~K ($T$ $\geq$ 60~K only for \textbf{H}$\|$[111] and \textbf{I}$\|$[110]) regardless of the orientation of current and magnetic field. This result was attributed as a proof of the absence of additional channel of scattering caused by stripes. We argue, that charge-transport anisotropy is originated in $R$$_x$Lu$_{1-x}$B$_{12}$ due to the anisotropy of electron-phonon scattering on the one hand and the effects of Fermi surface (FS) topology (at low temperatures) on the other.
\end{abstract}

\maketitle

\section*{1. Introduction}\label{Sec.1}
Nonmagnetic lutetium dodecaboride (LuB$_{12}$) with completely filled 4$f$ shell ($n_{4f}$ = 14) is the analog of lanthanum hexaboride (LaB$_6$) in the $R$B$_{12}$ family. LuB$_{12}$ is a metal \cite{1}-\cite{3} with small enough Sommerfeld coefficient $\gamma$ $\approx$ 3.3 $-$ 4.1~mJ/(mol$\cdot$K$^2$) \cite{4}-\cite{6} and one conduction electron per formula unit \cite{7}. LuB$_{12}$ becomes superconducting at $T_{\textmd{c}}$ $\sim$ 0.4 $-$ 0.44~K \cite{2}, \cite{8}-\cite{10}\footnotemark{}. \footnotetext{Superconductivity in LaB$_6$ is discussed briefly in \cite{10}.} Band structure calculations of lutetium dodecaboride suggest, that there are two conduction bands intersecting the Fermi level \cite{7, 11}. However, two-band superconductivity for LuB$_{12}$ has not detected \cite{9}. Low electronic work function of LuB$_{12}$ ($\phi$ $\approx$ 3.09 $-$ 3.18~eV at 1400~K) indicate on possible practical applications \cite{12}, although being higher in comparison with competitors among rich borides such as LaB$_6$ [$\phi$(100) $\approx$ 2.52~eV at 1600~K], CeB$_6$ [$\phi$(001) $\approx$ 2.62 eV at 1600~K] and GdB$_6$ ($\phi$ $\approx$ 1.53 eV for single-crystalline nanowire), \cite{13}-\cite{16}, see also \cite{17}.

Similar to other dodecaborides, LuB$_{12}$ crystallizes in the $fcc$ NaCl-type structure [sp. gr. $Fm$-3$m$-$O_h^5$ (No. 225)], Fig.\hyperref[FigX1]{1a}. In such lattice each rare earth (RE) ion is surrounded by truncated cuboctahedra formed by 24 boron atoms (B$_{24}$) with the radius $R$(B$_{24}$) $\sim$ 2.78~$\textmd{\AA}$ \cite{18}. As in the case of $R$B$_6$ class \cite{19, 20} the stability of $R$B$_{12}$ structure is realized among other things due to the discrepancy between inter- and intra-bond distances of boron atoms, which are $R$(B-B)$_{\textmd{inter}}$ $\approx$ 1.704~$\textmd{\AA}$ and $R$(B-B)$_{\textmd{intra}}$ $\approx$ 1.7874~$\textmd{\AA}$ for lutetium dodecaboride \cite{1}. Atomic/ionic radius of Lu$^{3+}$ ion [$R_\textmd{a}$(Lu) $\approx$ 1.737~$\textmd{\AA}$ and $R_i$(Lu$^{3+}$) $\approx$ 0.848~$\textmd{\AA}$], \cite{21, 22} as well as covalent boron radius $\approx$ 0.88~$\textmd{\AA}$ is much smaller compared to both the lattice constant $a$(LuB$_{12}$) $\approx$ 7.4648~$\textmd{\textmd{\AA}}$ \cite{1} and $R$(B$_{24}$). Such large difference between $R_i$ and $R$(B$_{24}$) results in a weak chemical bond of Lu-B with the formation of loosely bound Lu states or quasilocal mode (Einstein oscillator) detected at $\sim$ 13.8 $-$ 14.7~meV (160 $-$ 171~K) from various experiments including specific heat \cite{4, 6}, inelastic neutron scattering \cite{23, 24}, point-contact spectroscopy \cite{9}, transport \cite{25} and extended X-ray absorption fine structure (EXAFS) measurements \cite{26}. In \cite{27} it was proposed, that $R$B$_6$ (as well as $R$B$_{12}$) lattice may be generally divided into two sub-lattices: the rigid boron framework with Debye-type phonon spectrum and Einstein oscillators caused by RE ions. Therefore, such frame-clustered structures including dodecaborides as well as hexaborides may be considered as a model objects for charge transport and thermal investigations.

LuB$_{12}$ is often used in the literature as a reference system to extract magnetic contribution of various characteristics for the other members of $R$B$_{12}$ class. The application of this procedure may lead to incorrect results due to the noticeable renormalization of phonon spectra detected in the family of RE dodecaborides. For example, EXAFS experiments show the reduction of Einstein temperature in the set HoB$_{12}$ $-$ LuB$_{12}$ from maximal $\Theta_\textmd{E}$(HoB$_{12}$) $\approx$ 206~K (17.8~meV) down to minimal one $\Theta_\textmd{E}$(LuB$_{12}$) $\approx$ 160~K (13.8~meV), \cite{26}.

Here we report transport characteristics of $R$$_x$Lu$_{1-x}$B$_{12}$ ($R$$-$Dy, Er, Tm, Yb) substituted compounds with $x$(RE) $\leq$ 0.03. The aim of current work is not only to study the influence of magnetic impurity on the properties of LuB$_{12}$ host system, but also to compare the data obtained with the results published previously for LaB$_6$-based hexaborides with the formula $R$$_{0.01}$La$_{0.99}$B$_6$ ($R$$-$La, Ce, Pr, Nd, Eu, Gd, Ho) in \cite{25}. The investigation of the objects with so low doping levels is convenient way to elucidate the mechanisms responsible for significant transport anisotropy in LuB$_{12}$ nonmagnetic compound and to prove or exclude possible formation of charge stripes proposed in literature along the direction [110] below $\sim$ 150~K. Moreover, the research of Yb$_x$Lu$_{1-x}$B$_{12}$ set with isolated ytterbium ions may shed the light on the ground state formation in Kondo-isolator YbB$_{12}$.

The paper is organized as follows: experimental details and results are shown in Sections (Sects) \hyperref[Sec2]{2} and \hyperref[Sec3]{3}, respectively. In the discussion part \hyperref[Sec4p1]{4.1} a detailed analysis of zero-field resistivity is undertaken and several models are compared. In Sect.\hyperref[Sec4p2]{4.2} the role of Fermi surface effects is discussed and the anisotropy of electron-phonon scattering is proposed as alternative mechanism (Sect.\hyperref[Sec4p3]{4.3}) causing transport anisotropy instead charge stripes. A brief description of the results from the literature in favor of the applied approach is given in Sect.\hyperref[Sec4p4]{4.4}. Final conclusions are formulated in Sect.\hyperref[Sec5]{5}.

\section*{2. Materials and Methods}\label{Sec2}
\textbf{2.1}\label{Sec2p1} High quality single crystals of $R$$_{0.01}$Lu$_{0.99}$B$_{12}$ ($R$$-$Dy, Er, Tm, Yb) and $R$$_x$Lu$_{1-x}$B$_{12}$ (only for $R$-Tm and Yb with $x$ $\leq$ 0.03) series were prepared in IPM NASU (Kyiv, Ukraine) by using the induction crucible-free zone melting method in the sealed chamber of the 'Crystal-111A' setup \cite{17} in an argon gas atmosphere. The details about the synthesis of RE dodecaborides are described in \cite{17, 28}. Optimization of various technological parameters such as the rate of crystallization, the speed of rotation of the feed rod and the growing crystal, etc made it possible to grow perfect single crystals with a diameter of 5 $-$ 6~mm and length up to 80~mm [see the picture of the crystals in Figs.\hyperref[FigX1]{1g}, \hyperref[FigX1]{1i} and also in Fig.S1 in Supplementary Materials (SM)]. Electron microprobe analysis was used to estimate real concentration of isolated RE impurity ($x$$_\textmd{r}$), which is ranging from 0.9~$\%$ up to 3.5~$\%$ (Table \hyperref[Tab1]{1}). The difference in the Lu/$R$ ratio between the nominal compositions of the initial sintered rods and the real compositions of the grown crystals is caused by both the difference in the vapor pressure of lutetium and alloying RE elements and also by the difference in their distribution coefficients in the melt. For convenience all objects will be further designated by their nominal compositions. The residual resistivity ratio of the crystals RRR = $\rho$(300~K)/$\rho$$_0$ changes in the interval 2.5 $-$ 68 (Table \hyperref[Tab1]{1}).

The samples quality and orientation was controlled by electron microscopy and the X-ray methods. In particular, we performed X-ray topographic selective study of the quality of LuB$_{12}$ single crystal grown in the direction [001], (Fig.\hyperref[FigX1]{1b}). The major part of the LuB$_{12}$ [001] crystal rod ({\O} 5 $-$ 6~mm) consists of a single-crystalline core, that is free of grain boundaries and slip bands, in accordance with the Laue backscattering patterns, which is surrounded by a thin ring ($\leq$ 0.2~mm) with a small number of subgrains, misoriented relative to the core by no more than 5$^{\circ}$. Electron diffraction patterns (Fig.\hyperref[FigX1]{1c} and Figs.S2a, S2b in SM) and symmetric electron Kikuchi patterns (Fig.\hyperref[FigX1]{1e} and Fig.S2c in SM) correspond only to UB$_{12}$ structural type. X-ray diffraction pattern (Fig.\hyperref[FigX1]{1d}), and Laue backscattering patterns (Figs.\hyperref[FigX1]{1f}, \hyperref[FigX1]{1h} and Figs.S3a$-$S3c in SM), also prove high quality of the crystals under investigation. More details about the characterization of the samples are given in SM.

Rectangular bars for galvanomagnetic measurements (Figs.\hyperref[FigX1]{1g}, \hyperref[FigX1]{1i}) were cut by electric spark method from appropriate plates of initial single crystals (see Fig.S3d in SM), oriented with the help of X-ray diffractometer. Each object examined here was prepared in experimental geometry \textbf{I}$\|$[110] and \textbf{H}$\|$[110], (\textbf{I} $\perp$ \textbf{H}). Additional measurements were performed for Tm$_{0.03}$Lu$_{0.97}$B$_{12}$ and for Dy$_{0.01}$Lu$_{0.99}$B$_{12}$ compositions with different current (\textbf{I}$\|$[100], \textbf{I}$\|$[110], \textbf{I}$\|$[111]) and field (\textbf{H}$\|$[100], \textbf{H}$\|$[110], \textbf{H}$\|$[111]) geometries. To avoid deformation distortions of the surface layer, caused by electric spark cutting and polishing, all crystals after polishing were subjected to etching in dilute boiling nitric acid (HNO$_3$:H$_2$O mixture, 1:1).

\begin{figure*}[!b]
\begin{center}
\includegraphics[width = 17cm]{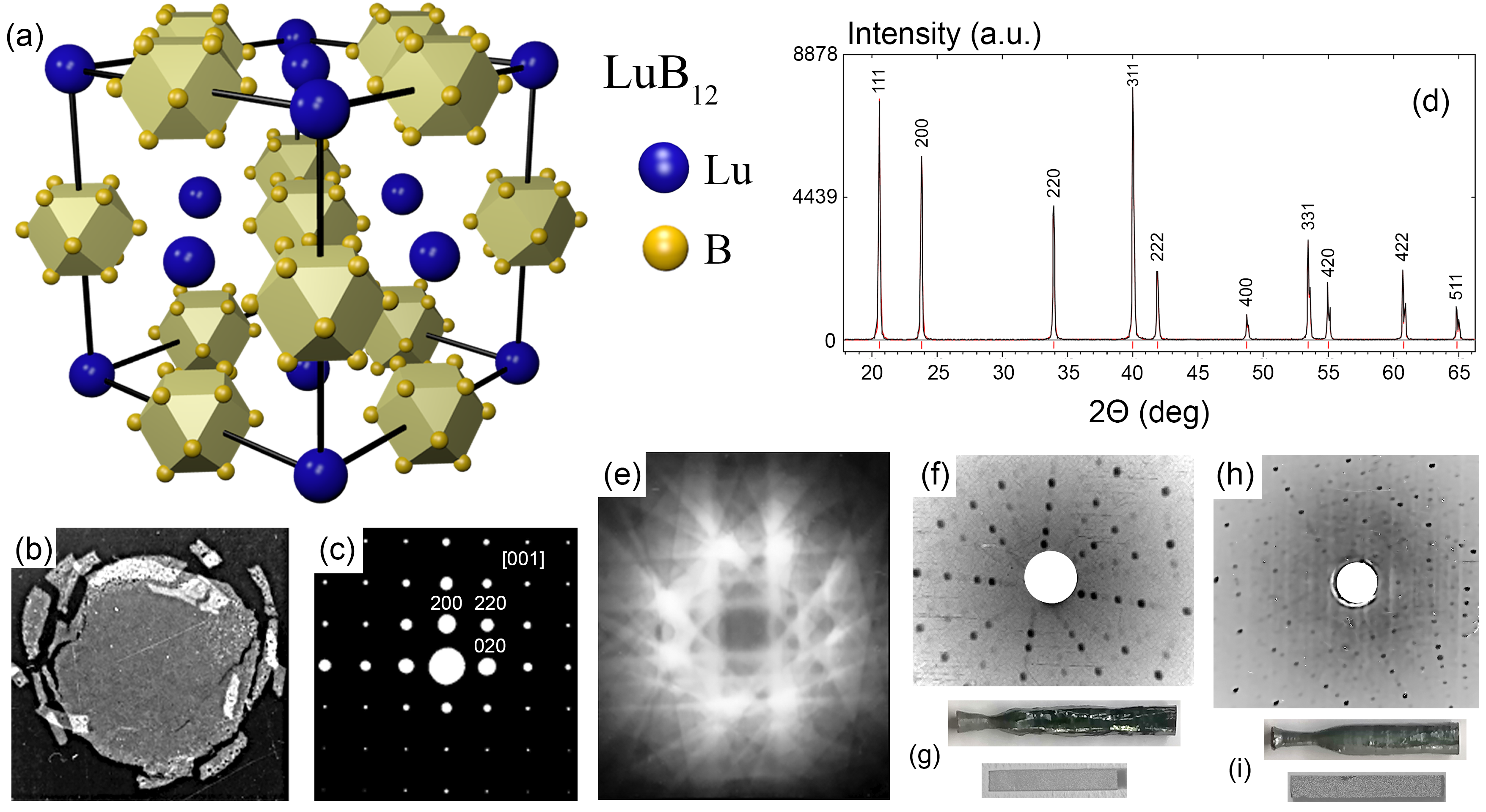}
   \parbox{17cm}{\caption{(Colour on-line). (a) Crystal structure of RE dodecaborides. (b) X-ray topogram from the lateral cross-section of as-grown LuB$_{12}$ single crystal with the growth direction [001]. (c) Electron diffraction pattern from local regions of LuB$_{12}$ single crystal along [001] axis. (d) Typical X-ray diffraction pattern in Cu K$_{\alpha}$ radiation with a Ni filter for crushed LuB$_{12}$ single crystal. (e) Symmetric electron Kikuchi pattern along the [001] direction in LuB$_{12}$ single crystal. The panels (f), (h) display Laue backscattering patterns and (g), (i) the pictures of several samples [including as-grown ones and the studied rectangular bars] of (f), (g) Tm$_{0.01}$Lu$_{0.99}$B$_{12}$; and (h), (i) Yb$_{0.03}$Lu$_{0.97}$B$_{12}$ compositions, respectively. Some additional figures with the characterization of the samples (Figs.S1$-$S3) are also presented in Sect.S1 of Supplementary Materials. }}\label{FigX1}

   \end{center}
\end{figure*}

Because of LuB$_{12}$ end-point compound is known to be sensitive to the variation of boron isotope composition \cite{6} as well as to the concentration of the impurities \cite{28} depending on the technological regime of sample preparation, we choose as reference the system Lu$^{\textmd{nat}}$B$_{12}$ with natural content of boron isotopes (18.83~$\%$ of $^{10}$B and 81.17~$\%$ of $^{11}$B). It exhibits minimal residual resistivity and maximal RRR = 68. The same crystal has been already investigated by our group including the study of transport (the sample $\#$1a in \cite{28}) and thermal properties \cite{6}. The substituted materials $R$$_x$Lu$_{1-x}$B$_{12}$ were additionally characterized by magnetic and thermal measurements. Some of them have been partially published in \cite{30}.

\textbf{2.2.}\label{Sec2p2} Transport measurements were carried out by conventional four-probe $dc$-technique with a direct current commutation at temperatures 1.8 $-$ 300~K in applied magnetic fields up to 82~kOe.
All contacts of thin copper wires were fixed to the surface of the sample by electrical discharge, whereas the current ones were covered additionally by a tiny amount of silver paint. Symmetrical arrangement of voltage contacts was checked by using an MBS-10 optical microscope. Angular dependencies of resistivity and Hall resistivity presented here as additional data were measured in five-terminal scheme by using original stepwise sample rotation technique. Step motor allowed a step-by-step rotating of the sample ($\Delta$$\varphi$ = 1.8$^{\circ}$) fixing its position in a steady magnetic field (see the inset in Fig.\hyperref[FigX7]{7b}). In such experimental geometry vector \textbf{H} was applied perpendicular to the measuring current \textbf{I}$\|$[110] and the angle $\varphi$ = $\textbf{n}$ $^{\wedge}$ $\textbf{H}$ was varied in the range $\varphi$ = 0 $-$ 360$^{\circ}$. In this study special attention was paid to high accuracy of temperature and magnetic field stabilization ($\Delta$$T$ $\approx$ 2~mK at $T$ $<$ 30~K and $\Delta$$H$ $\approx$ 2~Oe). Both parameters were achieved by using original commercial temperature controller TC-1.5/300 (Cryotel Ltd.) and superconducting magnet power supply unit SMPS-100 (Cryotel Ltd.) in combination with CERNOX-1050 thermometers (Lake Shore Cryotronics Inc.) and Hall sensors (Sensor RPC), correspondingly.

\newpage
\renewcommand{\thetable}{\arabic{table}}
\setcounter{table}{0}
\LTcapwidth=16cm
\begin{longtable*}[b]{cccccccccccc}
\caption{Analysis summary for $R$$_x$Lu$_{1-x}$B$_{12}$ materials with $x$ $\leq$ 0.03 ($R$-Dy, Er, Tm, Yb and Lu) by Eqs.(\hyperref[Eq.1]{1})-(\hyperref[Eq.3]{3}) ($N$-processes only): $x_{\textmd{r}}$ is the real concentration of RE impurity; RRR and $\rho_0$ are residual resistivity ratio RRR = $\rho$(300~K)/$\rho_0$ and residual resistivity; $\lambda_{\textmd{tr}}$$\omega_\textmd{D}$/$\omega_\textmd{P}$$^2$ and ($KN$/$m$)$_i$ are amplitude factors in Eq.(\hyperref[Eq.2]{2}) and Eq.(\hyperref[Eq.3]{3}), respectively; $\Theta_{\textmd{D}}$ and $\Theta_{\textmd{Ei}}$ are Debye and Einstein temperatures; MR$_+$ and A$_{\textmd{MR}}^{\textmd{max}}$ are the amplitude of positive magnetoresistance and the amplitude of the anisotropy of magnetoresistance at fixed magnetic field $H$ = 80~kOe (experimental geometry \textbf{I}$\|$[110], \textbf{H}$\|$[110]). The data for Lu$^{\textmd{nat}}$B$_{12}$ and Ho$_{0.01}$Lu$_{0.99}$B$_{12}$ crystals were taken from previous works of our group \cite{28} and \cite{31}, correspondingly.}\label{Tab1}\\
\hhline{============} \\
\quad\quad \multirow{3}{*}{$R$$_x$Lu$_{1-x}$B$_{12}$}	\quad& \quad\quad $x_{\textmd{r}}$	& \multirow{3}{*}{RRR}     & $\rho_0$ & ($KN$/$m$)$_1$ & $\Theta_{\textmd{E1}}$ & ($KN$/$m$)$_2$ & $\Theta_{\textmd{E2}}$ & $\lambda_{\textmd{tr}}$$\omega_\textmd{D}$/$\omega_\textmd{P}$$^2$ & $\Theta_{\textmd{D}}$ & \quad MR$_{+}$ \quad& A$_{\textmd{MR}}^{\textmd{max}}$ \quad\quad \\
 \\
   \quad\quad	\quad& \quad\quad($\%$)  &    &    ($\mu$$\Omega$$\cdot$cm)  & (m$\Omega$$\cdot$cm$\cdot$K)  &  (K) & (m$\Omega$$\cdot$cm$\cdot$K) & (K) & ($\mu$$\Omega$$\cdot$cm) & (K) & \quad($\%$)  \quad & ($\%$) \quad\quad\\
    \\
\hhline{------------} \\
			\quad\quad Lu$^{\textmd{nat}}$B$_{12}$  \quad&  \quad\quad $-$  &  68.1  &    0.16  & 0.157  &  150 & 2.52 & 365 & 0.455 & 1160 & \quad 575 \quad & 480 \quad\quad\\
			 \\
				\quad\quad Dy$_{0.01}$Lu$_{0.99}$B$_{12}$  \quad&  \quad\quad 0.9  &  35  &    0.31  & 0.21 &  154 & 2.45 & 361 & 0.345 & 1160 & 72$-$188 & 114 \quad\quad \\
			 \\
			\quad\quad	Er$_{0.01}$Lu$_{0.99}$B$_{12}$  \quad&  \quad\quad 1.3  &  25  &    0.43  & 0.205  &  156 & 2.25 & 361 & 0.305 & 1160 & \quad 50 \quad & 46.3 \quad\quad \\
			 \\
			\quad\quad	Ho$_{0.01}$Lu$_{0.99}$B$_{12}$  \quad&  \quad\quad 1  &  15.7  &    0.72  & 0.222  &  153 & 2.26 & 357 & 0.347 & 1160 & \quad 26 \quad & $-$ \quad\quad \\
			  \\
			\quad\quad	Tm$_{0.01}$Lu$_{0.99}$B$_{12}$  \quad&  \quad\quad 1.5  &  27.4  &    0.39  & 0.191  &  153 & 2.61 & 367 & 0.31 & 1160 & \quad 58 \quad & 59.7 \quad\quad \\
			  \\
			\quad\quad	Yb$_{0.01}$Lu$_{0.99}$B$_{12}$  \quad&  \quad\quad 1.2  &  6.5  &    0.18  & 0.167  &  154 & 2.6 & 357 & 0.305 & 1160 & \quad 9.7 \quad & 0.7 \quad\quad \\
\\
\hhline{------------} \\
			 \quad\quad	Tm$_{0.03}$Lu$_{0.97}$B$_{12}$  \quad& \quad\quad 1.8 &  26.7  &   0.42  & 0.225  & 154  & 2.34  & 357 & 0.336 & 1160 & \quad 57$-$112 \quad & 60.4   \quad\quad \\
			 \\
		\quad\quad	Yb$_{0.02}$Lu$_{0.98}$B$_{12}$  \quad& \quad\quad 2  & 4.7  &    3.0 & 0.188  & 164 & 2.52  & 362 & 0.306 & 1160 & \quad 3.5 \quad & $-$ \quad\quad \\
			  \\
		\quad\quad		Yb$_{0.03}$Lu$_{0.97}$B$_{12}$  \quad&  \quad\quad 3.5   &  2.5 &   6.3 & 0.165 &  172  & 2.95 & 360 & 0.331 & 1160 & \quad 1 \quad & $-$ \quad\quad \\
\\
		\hhline{============}
\end{longtable*}

\noindent

\section*{3. Results}\label{Sec3}

\begin{figure*}[!h]
\includegraphics[width = 18cm]{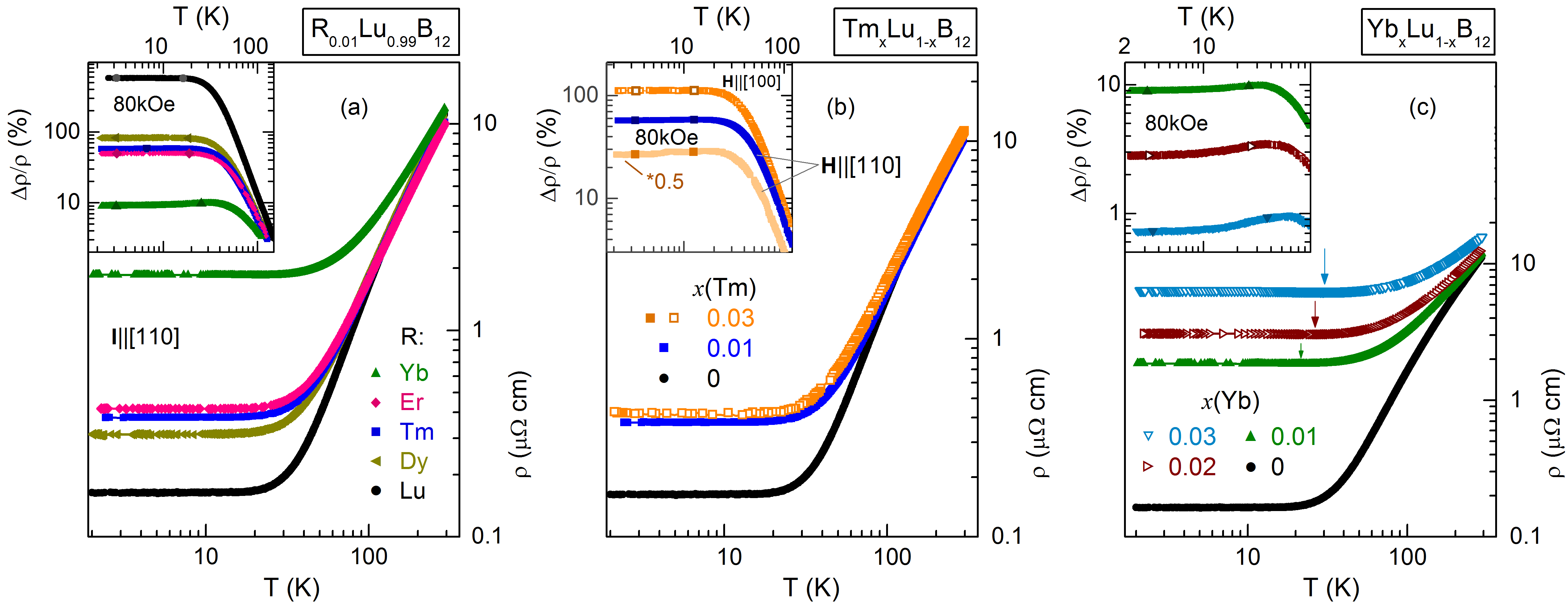}
   \parbox{18cm}{\caption{(Colour on-line). Temperature evolution of zero-field electrical resistivity $\rho$($T$) presented for (a) $R$$_{0.01}$Lu$_{0.99}$B$_{12}$ ($R$-Dy, Er, Tm, Yb, and Lu) series, and separately for diluted systems (b) Tm$_{x}$Lu$_{1-x}$B$_{12}$, and (c) Yb$_{x}$Lu$_{1-x}$B$_{12}$ with $x$(Tm, Yb) $\leq$ 0.03. Real content of isolated RE impurities ($x_\textmd{r}$) is given in Table \hyperref[Tab1]{1}. The arrows on panel (c) designate the position of the minimum $T_{\textmd{min}}$. The insets on each panel display the $T$-dependencies of transverse magnetoresistance $\Delta\rho$($T$)/$\rho$ measured in experimental geometry \textbf{I}$\|$[110], \textbf{H}$\|$[110]  at fixed point $H$ = 80~kOe. MR data for Tm$_{0.03}$Lu$_{0.97}$B$_{12}$ are presented for two orientations of magnetic field \textbf{H}$\|$[100] and \textbf{H}$\|$[110] (and the same current geometry \textbf{I}$\|$[110]). For the last one the curve $\Delta\rho$($T$)/$\rho$ is shifted along vertical axis by the factor 0.5 for convenience.}}\label{FigX2}
\end{figure*}

\begin{figure}[!htpb]
\begin{center}
\includegraphics[width = 8 cm]{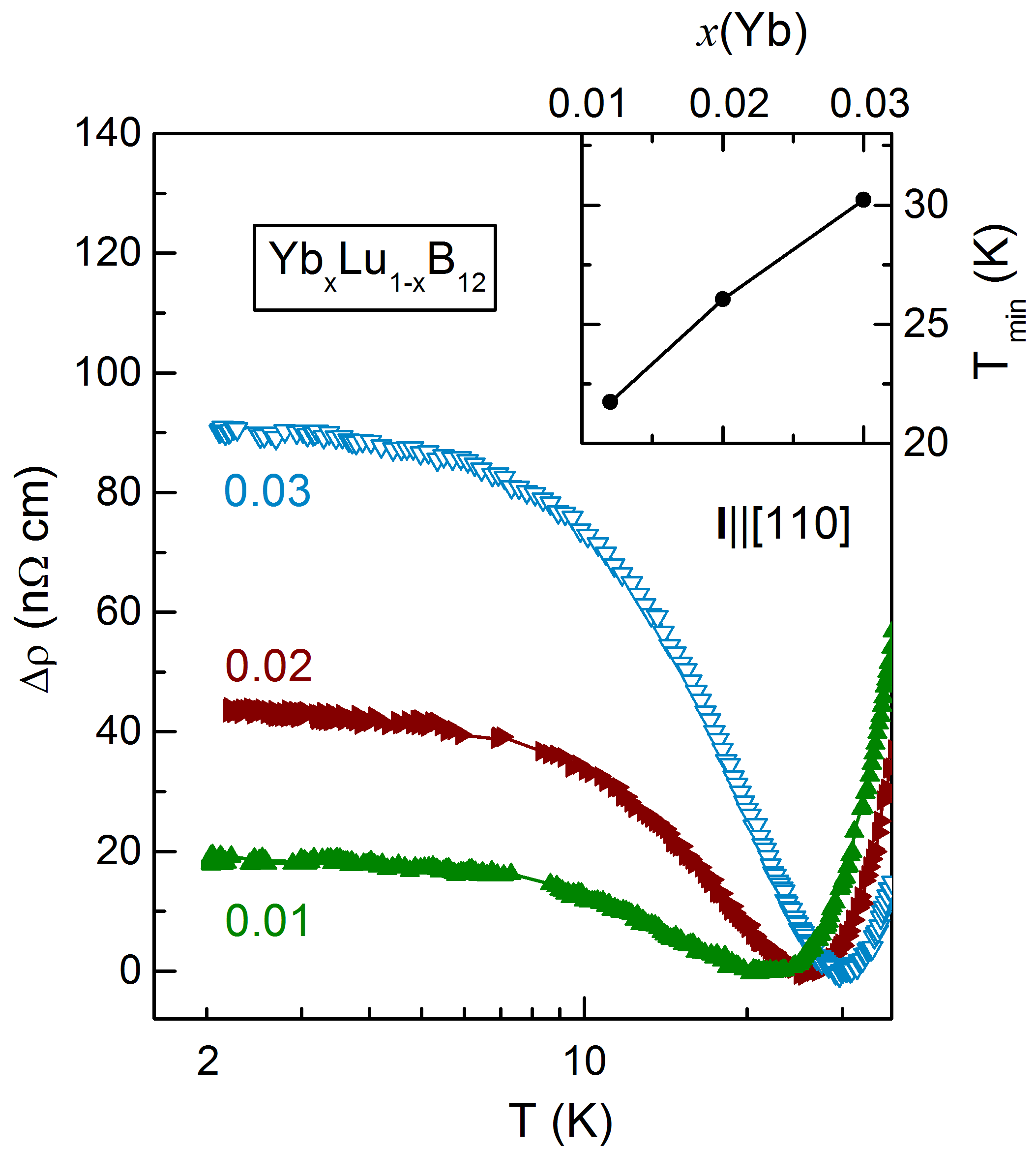}
  \caption{(Colour on-line). Low-temperature rise of resistivity $\Delta\rho$($T$) = $\rho$($T$) $-$ $\rho_0$ vs $T$ for Yb$_{x}$Lu$_{1-x}$B$_{12}$ series in enlarged scale. Current geometry was \textbf{I}$\|$[110]. The values of residual resistivity $\rho_0$ are given in Table  \hyperref[Tab1]{1}. The inset displays how the position of  $T_\textmd{{min}}$ varies with real concentration of Yb.}\label{FigX3}
   \end{center}
\end{figure}

\textbf{3.1}\label{Sec3p1} Zero-field electrical resistivity of $R$$_{0.01}$Lu$_{0.99}$B$_{12}$ solid solutions ($R$$-$Dy, Er, Tm, Yb, and Lu) is demonstrated as a function of temperature in Fig.\hyperref[FigX2]{2a}. For convenience we present separately the $\rho$($T$) data of Tm$_x$Lu$_{1-x}$B$_{12}$ and Yb$_x$Lu$_{1-x}$B$_{12}$ materials in Fig.\hyperref[FigX2]{2b} and Fig.\hyperref[FigX2]{2c}, respectively. Here the data for current geometry \textbf{I}$\|$[110] are shown. The curves measured for all the compounds under investigation exhibit typical metallic behavior with different regimes at low-$T$ limit. In particular, ($i$) temperature independent 'plateau' ($\rho$$_0$) appears below 20 $-$ 25~K for LuB$_{12}$ and also for Dy$_{0.01}$Lu$_{0.99}$B$_{12}$, Er$_{0.01}$Lu$_{0.99}$B$_{12}$, Tm$_{0.01}$Lu$_{0.99}$B$_{12}$, and Tm$_{0.03}$Lu$_{0.97}$B$_{12}$ diluted compounds. ($ii$) When Lutetium is replaced by ytterbium in Yb$_x$Lu$_{1-x}$B$_{12}$, the $\rho$($T$) curve passes through the minimum at $T_{\textmd{min}}$ with following rise by the magnitude $\Delta$$\rho$($T$) = $\rho$($T$) $-$ $\rho$$_0$ $\leq$ 0.1~$\mu\Omega\cdot$cm \footnotemark{}  and with an evident tendency to saturation in the interval $T$ $<$ 32~K as shown on enlarged scale in Fig.\hyperref[FigX3]{3}.
\footnotetext{$\rho$$_0$ designates the value of resistivity at the minimal point. It is considered here as residual resistivity (see Sect.\hyperref[Sec4p1]{4.1}).} Such dependencies are typical for Kondo impurity behavior in normal metal. The position of $T_{\textmd{min}}$ increases with Yb-doping (inset in Fig.\hyperref[FigX3]{3}). Residual resistivity is a minimal $\rho$$_0$ $\approx$ 0.16~$\mu\Omega\cdot$cm (RRR = 68) in Lu$^{\textmd{nat}}$B$_{12}$, and it shifts up for substituted materials achieving a maximal value $\rho$$_0$ $\approx$ 6.28~$\mu\Omega\cdot$cm (RRR = 2.5) in Yb$_{0.03}$Lu$_{0.97}$B$_{12}$ (see Fig.\hyperref[FigX2]{2c} and Table \hyperref[Tab1]{1}). For comparison the data of Ho$_{0.01}$Lu$_{0.99}$B$_{12}$ were taken from previous work of our group \cite{31} and presented in Tables \hyperref[Tab1]{1} - \hyperref[Tab2]{2} and also in Figs.S4a, S5 in Supplementary Materials. Note also that the influence of current geometry on zero-field resistivity was checked and not revealed for Tm$_{0.03}$Lu$_{0.97}$B$_{12}$ (Fig.S4b in SM).

\begin{figure*}[!t]
\includegraphics[width = 18cm]{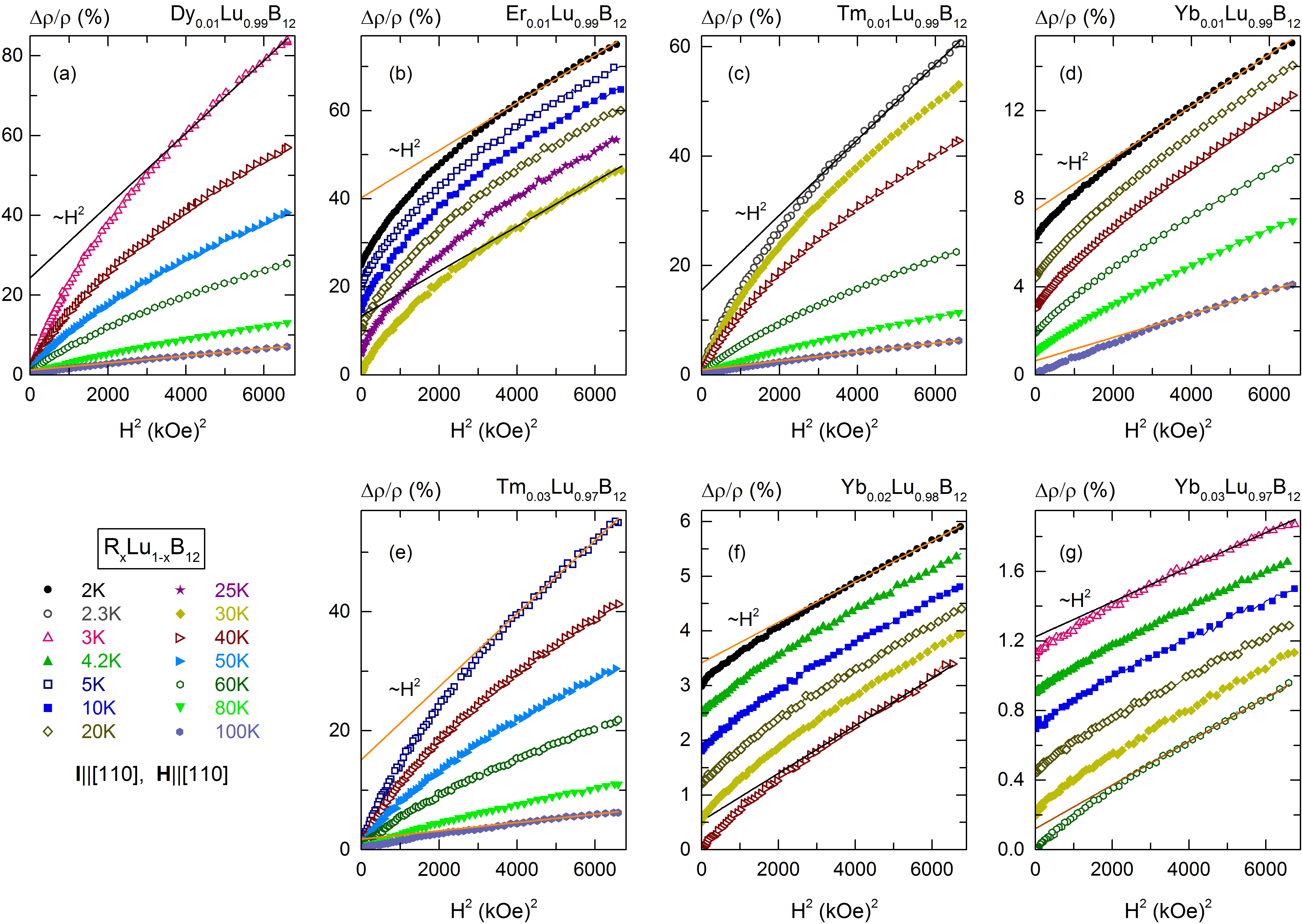}
   \parbox{18cm}{\caption{(Colour on-line). Field scans of transverse magnetoresistance $\Delta\rho$($H$)/$\rho$ = $f$($H^2$, $T$) of $R$$_x$Lu$_{1-x}$B$_{12}$ materials at various fixed temperatures in the range 2 $-$ 100~K. Experimental geometry was \textbf{I}$\|$[110] and \textbf{H}$\|$[110]. Solid lines display quadratic asymptotic $\sim$ $H^2$. The isotherms on panels (b), (d), (f), and (g) are shifted along the vertical axis for convenience.}}\label{FigX4}
\end{figure*}

\textbf{3.2}\label{Sec3p2} To illustrate the evolution of transverse magnetoresistance (MR) we measured both temperature (the insets in Figs.\hyperref[FigX2]{2a}-\hyperref[FigX2]{2c}, and Fig.S4 in SM) and field (Fig.\hyperref[FigX4]{4} and Figs.S5$-$S8 in SM) scans. The main results presented in this work were obtained in experimental geometry \textbf{I}$\|$[110] and \textbf{H}$\|$[110]. According to Fig.\hyperref[FigX2]{2} all the compounds under investigation are characterized by positive magnetoresistance (MR$_{+}$), which amplitude changes by more than two orders [$\Delta\rho$/$\rho$($H$ = 80~kOe) $\approx$ 1 $-$ 575~$\%$, Table \hyperref[Tab1]{1}]. Note that, MR$_{+}$ of Yb$_x$Lu$_{1-x}$B$_{12}$ materials falls the most considerably [at the maximum point $\Delta\rho$/$\rho$($H$ = 80~kOe) $\approx$ 1 $-$ 9.7~$\%$, see the inset in Fig.\hyperref[FigX2]{2c}]. Such effect allows to propose a crossover from positive to negative MR regime in Yb$_x$Lu$_{1-x}$B$_{12}$ series, when the concentration of ytterbium exceeds some threshold value in $\Delta$$x_+$ vicinity of $x$(Yb) $\approx$ 0.03.

Fig.\hyperref[FigX4]{4} displays field dependencies of magnetoresistance in coordinates $\Delta\rho$/$\rho$ = $f$($H$$^2$, $T_0$) measured for \hyperref[FigX4]{(a)}-\hyperref[FigX4]{(d)} $R$$_{0.01}$Lu$_{0.99}$B$_{12}$ family and also for \hyperref[FigX4]{(e)} Tm$_{0.03}$Lu$_{0.97}$B$_{12}$, \hyperref[FigX4]{(f)} Yb$_{0.02}$Lu$_{0.98}$B$_{12}$, \hyperref[FigX4]{(g)} Yb$_{0.03}$Lu$_{0.97}$B$_{12}$ compositions. (For convenience the isotherms on panels \hyperref[FigX4]{(b)}, \hyperref[FigX4]{(d)}, \hyperref[FigX4]{(f)}, and \hyperref[FigX4]{(g)} are shifted along the vertical axis). Complete set of experimental data is presented for Tm$_{0.01}$Lu$_{0.99}$B$_{12}$, Yb$_{0.01}$Lu$_{0.99}$B$_{12}$, Dy$_{0.01}$Lu$_{0.99}$B$_{12}$ and Tm$_{0.03}$Lu$_{0.97}$B$_{12}$ in Fig.S7 in SM. From Fig.\hyperref[FigX4]{4} and Figs.S5$-$S7 in SM it is visible, that the saturation is not fully reached up to 82~kOe for geometry \textbf{H}$\|$[110] (\textbf{I}$\|$[110]). The variation of RE impurity as well as its concentration in LuB$_{12}$ end-point system changes only the amplitude of MR, while the field asymptotic is described by positive quadratic component $\Delta\rho$/$\rho$($H$) = $\mu$$_\textmd{D}^2$$H^2$ observed in the range 40/50 $-$ 82~kOe for all the compounds under investigation (Fig.\hyperref[FigX4]{4}) and also for Ho$_{0.01}$Lu$_{0.99}$B$_{12}$ composition (Fig.S5 in SM). Here, the coefficient $\mu$$_\textmd{D}$ may be considered as reduced drift mobility of the charge carriers (see Sect.\hyperref[Sec4p3]{4.3}). Note, that the anomalies caused by quantum oscillations (Shubnikov-de Haas  effect) were not found on the resistivity curves below 82~kOe (Figs.\hyperref[FigX4]{4} and S6, S8 in SM). According to \cite{28} weak quantum oscillations have been detected in magnetoresistance and Hall resistivity on the same Lu$^\textmd{{nat}}$B$_{12}$ crystal in experimental geometry \textbf{I}$\|$[110] and \textbf{H}$\|$[001] only above 100~kOe.

\begin{figure*}[!htpb]
\hspace{-1.5cm}\includegraphics[width = 19cm]{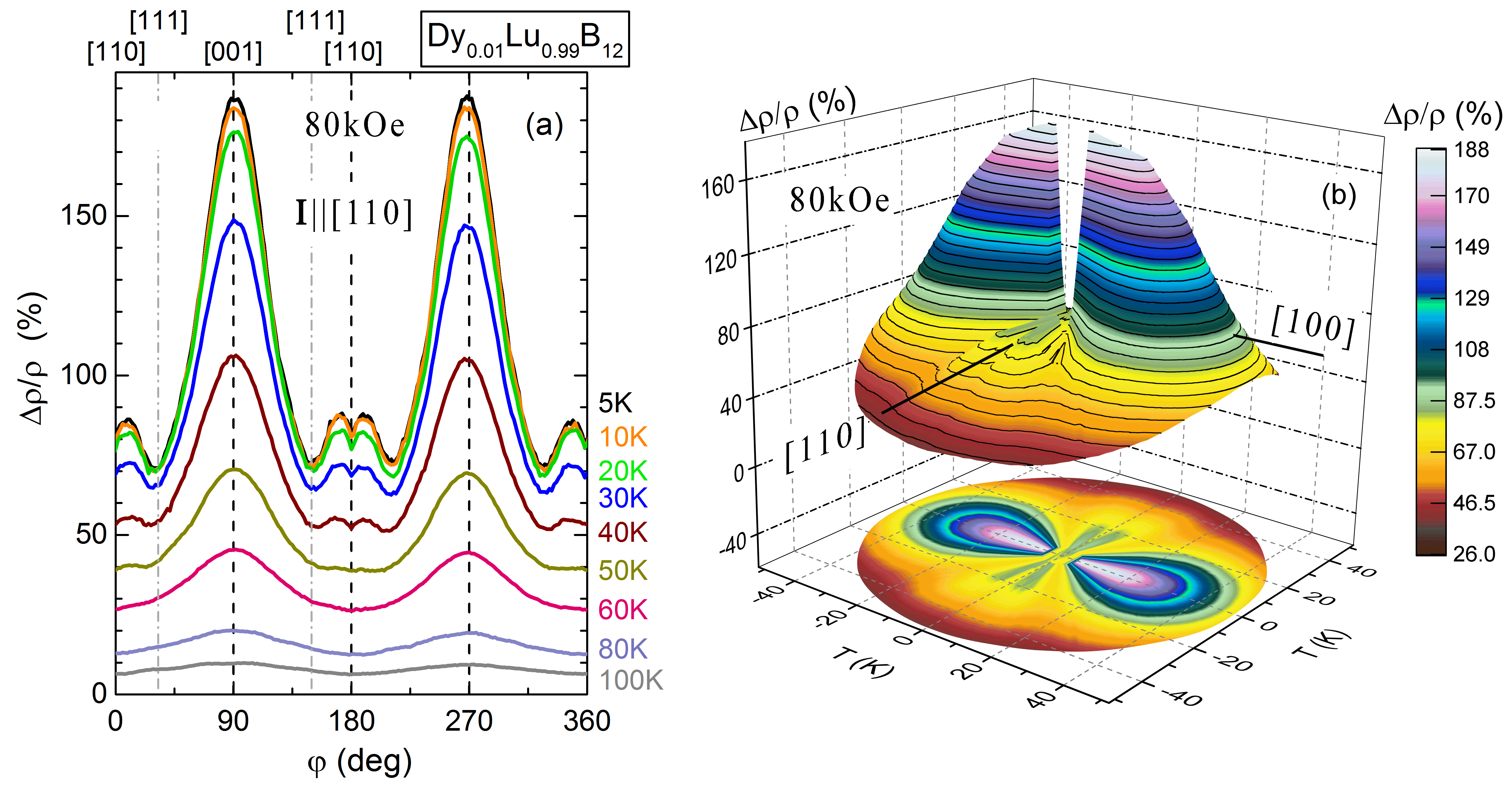}
   \parbox{18cm}{\caption{(Colour on-line). (a) Angular dependencies of transverse magnetoresistance for Dy$_{0.01}$Lu$_{0.99}$B$_{12}$ crystal at constant magnetic field $H$ = 80~kOe. The rotation was performed around \textbf{I}$\|$[110] axis. Vertical lines show the positions, when vector \textbf{H} is aligned with principal directions. (b) Presentation of MR of Dy$_{0.01}$Lu$_{0.99}$B$_{12}$ in cylindrical coordinates reconstructed by using the data from the panel (a) below 60~K. The orientations of \textbf{H} along [110] and [100] are displayed by solid lines. For comparison we also show the projection of 3D graph onto ($xy$) plane (color gradient polar plot). }}\label{FigX5}
\end{figure*}

\begin{figure}[htpb]
\begin{center}
\includegraphics[width = 14 cm]{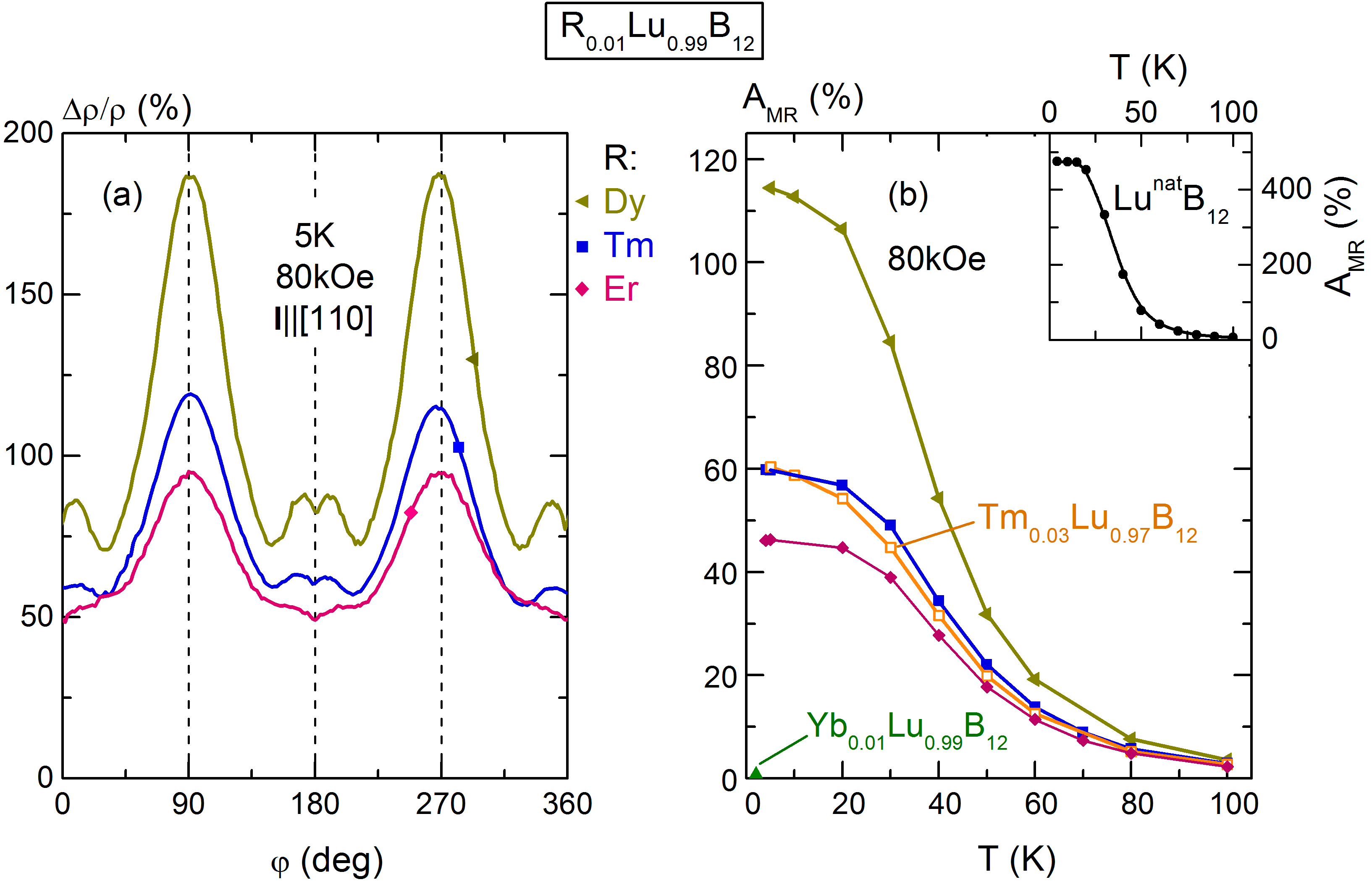}
  \caption{(Colour on-line). (a) The comparison of angular dependencies of transverse magnetoresistance of $R$$_{0.01}$Lu$_{0.99}$B$_{12}$ ($R$-Dy, Er, Tm, Yb) materials at the point $T$ = 5~K, $H$ = 80~kOe. The rotation was performed around \textbf{I}$\|$[110] axis. (b) Temperature evolution of the amplitude of MR$_+$ anisotropy $A_{\textmd{MR}}$($T$) calculated for $R$$_{0.01}$Lu$_{0.99}$B$_{12}$ ($R$-Dy, Er, Tm, Yb) family and also for Tm$_{0.03}$Lu$_{0.97}$B$_{12}$ composition at constant magnetic field $H$ = 80~kOe.   The inset shows the parameter $A_{\textmd{MR}}$($T$) for Lu$^{\textmd{nat}}$B$_{12}$ host system (the data were taken from \cite{28}).}\label{FigX6}
   \end{center}
\end{figure}

\begin{figure*}[!htpb]
\hspace{-1cm}\includegraphics[width = 18.8cm]{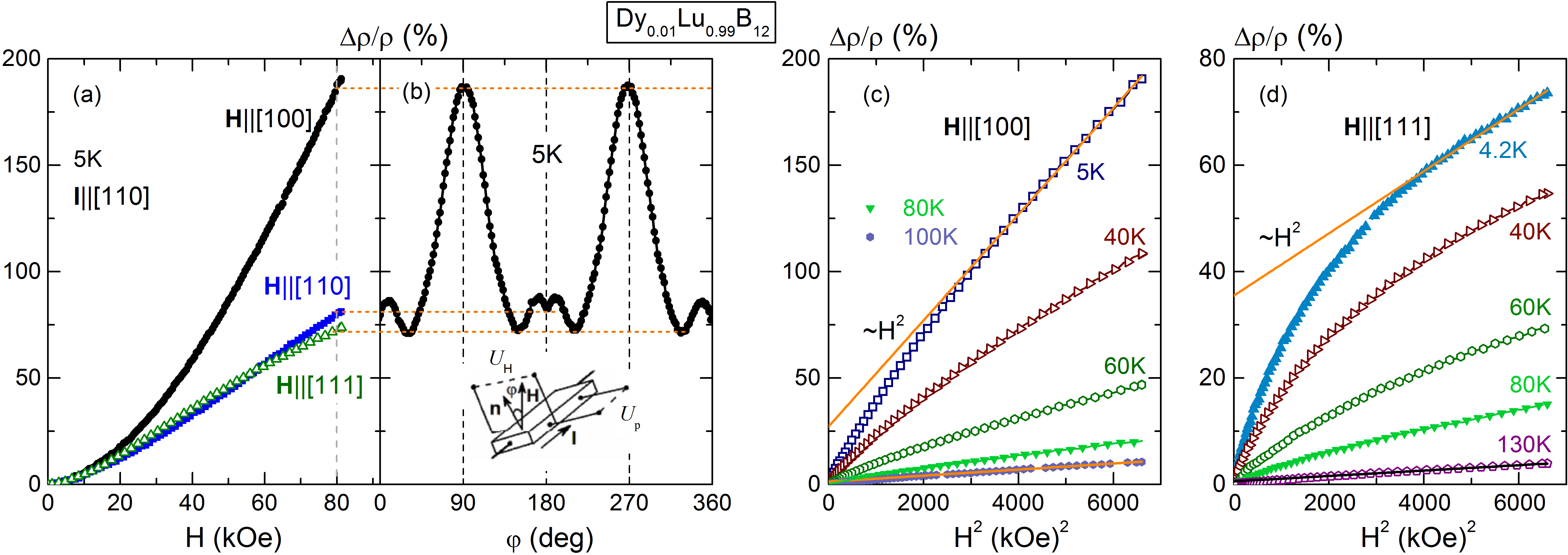}
   \parbox{18cm}{\caption{(Colour on-line). (a) Magnetic field scans and (b) angular dependencies of MR of Dy$_{0.01}$Lu$_{0.99}$B$_{12}$ sample at $T$ = 5~K. The inset in panel (b) shows the schematic view of the sample rotation experiment. \textbf{I} designates a measuring current, \textbf{n} $-$ the normal vector to the sample surface, the angle $\varphi$ = \textbf{n} $^{\wedge}$ \textbf{H}, both $U_\textmd{H}$ and $U_\textmd{p}$ are voltages from the Hall and potential probes to the crystal. (c)-(d) Field scans of transverse magnetoresistance $\Delta\rho$/$\rho$ = $f$($H^2$, $T$) of Dy$_{0.01}$Lu$_{0.99}$B$_{12}$ at various fixed temperatures from the range 4.2 $-$ 130~K. Experimental geometry was (c) \textbf{H}$\|$[100] and (d) \textbf{H}$\|$[111] with the same current orientation \textbf{I}$\|$[110]. Solid lines in (c) and (d) panels display quadratic asymptotic $\sim$ $H^2$. }}\label{FigX7}
\end{figure*}

\textbf{3.3}\label{Sec3p3} In order to prove or exclude the existence of additional channel of scattering caused by probable presence of dynamic charge stripes \cite{28} we performed the study of the anisotropy of magnetoresistance. As an objects of investigation we took Dy$_{0.01}$Lu$_{0.99}$B$_{12}$ and Tm$_{0.03}$Lu$_{0.97}$B$_{12}$ compounds (RRR = 35, 26.7, Table \hyperref[Tab1]{1}) with the minimal after host Lu$^{\textmd{nat}}$B$_{12}$ (RRR = 68) residual resistivity (Dy$_{0.01}$Lu$_{0.99}$B$_{12}$), since, as it was shown in \cite{28}, the increasing of $\rho_0$ in the set of lutetium dodecaboride single crystals leads to the depression of characteristic features on angular dependencies of MR even at helium temperature. The obtained data of angle sweeping and field measurements are illustrated in Figs.\hyperref[FigX5]{5}-\hyperref[FigX6]{6} and \hyperref[FigX7]{7}, respectively. To shed more light on the evolution of MR$_+$ anisotropy in Dy$_{0.01}$Lu$_{0.99}$B$_{12}$ the angular dependencies $\Delta\rho$/$\rho$ = $f$($\varphi$) were investigated at constant magnetic field $H$ = 80~kOe for the isotherms from the range $T$ = 4.2 $-$ 100~K (Fig.\hyperref[FigX5]{5}). A similar experiment was performed for the rest systems in $R$$_{0.01}$Lu$_{0.99}$B$_{12}$ set. However, for ease of perception the data obtained at the only one point $T$ = 5~K are presented here for comparison (Fig.\hyperref[FigX6]{6a}). According to Figs.\hyperref[FigX5]{5a}, \hyperref[FigX6]{6a} a broad peak is observed in the vicinity of \textbf{H}$\|$[100] direction. At the same time we don't see any additional narrow dips reported at \textbf{H}$\|$[100] of Lu$^{\textmd{nat}}$B$_{12}$ at low temperatures \cite{28}. For the other two directions of magnetic field \textbf{H}$\|$[110] and \textbf{H}$\|$[111] a more complicated $T$-behavior is registered. In particular, $\Delta\rho$/$\rho$ = $f$($\varphi$) curves change the form and instead a wide minimum along \textbf{H}$\|$[110] orientation another feature originates below 50~K (Fig.\hyperref[FigX5]{5a}). Accordingly minimal point is recorded now along \textbf{H}$\|$[111] direction, while a beak-shaped singular anomaly appears in a narrow interval of angles near \textbf{H}$\|$[110]. To visualize the evolution of the anisotropy of magnetoresistance we also present 3D view of MR$_+$ in the cylindrical coordinates together with its projection (color gradient polar plot), Fig.\hyperref[FigX5]{5b}. It is worth noting, that our data correlate well with the results reported earlier for Lu$^{\textmd{nat}}$B$_{12}$ in \cite{28} and also in \cite{3}.

Figs.\hyperref[FigX7]{7a}, \hyperref[FigX7]{7b} demonstrate comparison between field and angular dependencies of magnetoresistance of Dy$_{0.01}$Lu$_{0.99}$B$_{12}$ at the point $T$ = 5~K, while Figs.\hyperref[FigX7]{7c} and \hyperref[FigX7]{7d} display the family of curves $\Delta\rho$/$\rho$ = $f$($H$$^2$, $T_0$) for experimental geometries \textbf{H}$\|$[100] and \textbf{H}$\|$[111] (\textbf{I}$\|$[110]), respectively. As can be seen from Fig.\hyperref[FigX7]{7a}, unlimited growth of MR$_+$ is registered for orientations \textbf{H}$\|$[100] and \textbf{H}$\|$[110]. On the contrary, the data obtained for direction \textbf{H}$\|$[111] show a trend to saturation in the range $T$ $<$ 60~K having a different convexity (see also Fig.S7h in SM). In this case the saturation may be proposed above 80~kOe. Such result correlates with Fermi surface (FS) studies of LuB$_{12}$ \cite{3}, which predict the existence of only closed orbits in the vicinity of [111] (see the discussion in Sect.\hyperref[Sec4p2]{4.2}). According to Figs.\hyperref[FigX7]{7c}, \hyperref[FigX7]{7d} field scans presented in coordinates $\Delta\rho$/$\rho$ = $f$($H$$^2$, $T_0$) are also described by positive quadratic asymptotic $\Delta\rho$/$\rho$($H$) = $\mu$$_\textmd{D}^2$$H^2$ similar to the data from Fig.\hyperref[FigX4]{4}.
We measured the data for Tm$_{0.03}$Lu$_{0.97}$B$_{12}$ composition with different current geometries \textbf{I}$\|$[100] and \textbf{I}$\|$[111]  and the same field orientation \textbf{H}$\|$[110]. As can be seen from  Fig.S8 in SM the same behavior is observed.

The amplitude of the anisotropy of magnetoresistance defined as $A_{\textmd{MR}}$ = $\Delta\rho$/$\rho$$^{\textmd{max}}$($\varphi$) $-$ $\Delta\rho$/$\rho$$^{\textmd{min}}$($\varphi$) was also estimated for $R$$_{0.01}$Lu$_{0.99}$B$_{12}$ family at constant magnetic field $H$ = 80~kOe. Temperature evolution of $A_{\textmd{MR}}$  parameter is displayed in Fig.\hyperref[FigX6]{6b}. The function $A_{\textmd{MR}}$($T$) is rising with cooling achieving maximal values $A_{\textmd{MR}}^{\textmd{max}}$ $\approx$ 0.7 $-$ 114~$\%$ at low temperatures (Table \hyperref[Tab1]{1}). In  Lu$^{\textmd{nat}}$B$_{12}$ host system the parameter $A_{\textmd{MR}}^{\textmd{max}}$ reaches $\approx$ 480~$\%$ (the inset in Fig.\hyperref[FigX6]{6b}).

\section*{4. Discussion}\label{Sec4}
\subsection*{\emph{4.1. The analysis of zero-field resistivity.}}\label{Sec4p1}

The main difference between zero-field resistivity curves for $R$$_x$Lu$_{1-x}$B$_{12}$ ($x$ $\leq$ 0.03) set in comparison with similar data for $R$$_{0.01}$La$_{0.99}$B$_6$ ($R$$-$Ce, Pr, Nd, Ho) family is the absence of the region of low-temperature power-law growth as $\rho$($T$) $\sim$ $T^{-\alpha}$ with indexes $\alpha$ $\approx$ 0.2 $-$ 0.49 \cite{25}. When LaB$_6$ parent system is doped by small amount of magnetic impurity ($x_\textmd{r}$ $\leq$ 1.5~$\%$) this anomaly appears in the range 2 $-$ 20~K \cite{25}. Here, low-$T$ rise of $\rho$($T$) is detected only for Yb$_x$Lu$_{1-x}$B$_{12}$ family (Fig.\hyperref[FigX3]{3}). But both low amplitude of the effect [$\Delta$$\rho$($T$) $\leq$ 0.1~$\mu\Omega\cdot$cm] and rather narrow interval of its occurrence do not allow us to interpret it in the same way.

\begin{figure}[!t]
\begin{center}
\includegraphics[width = 14cm]{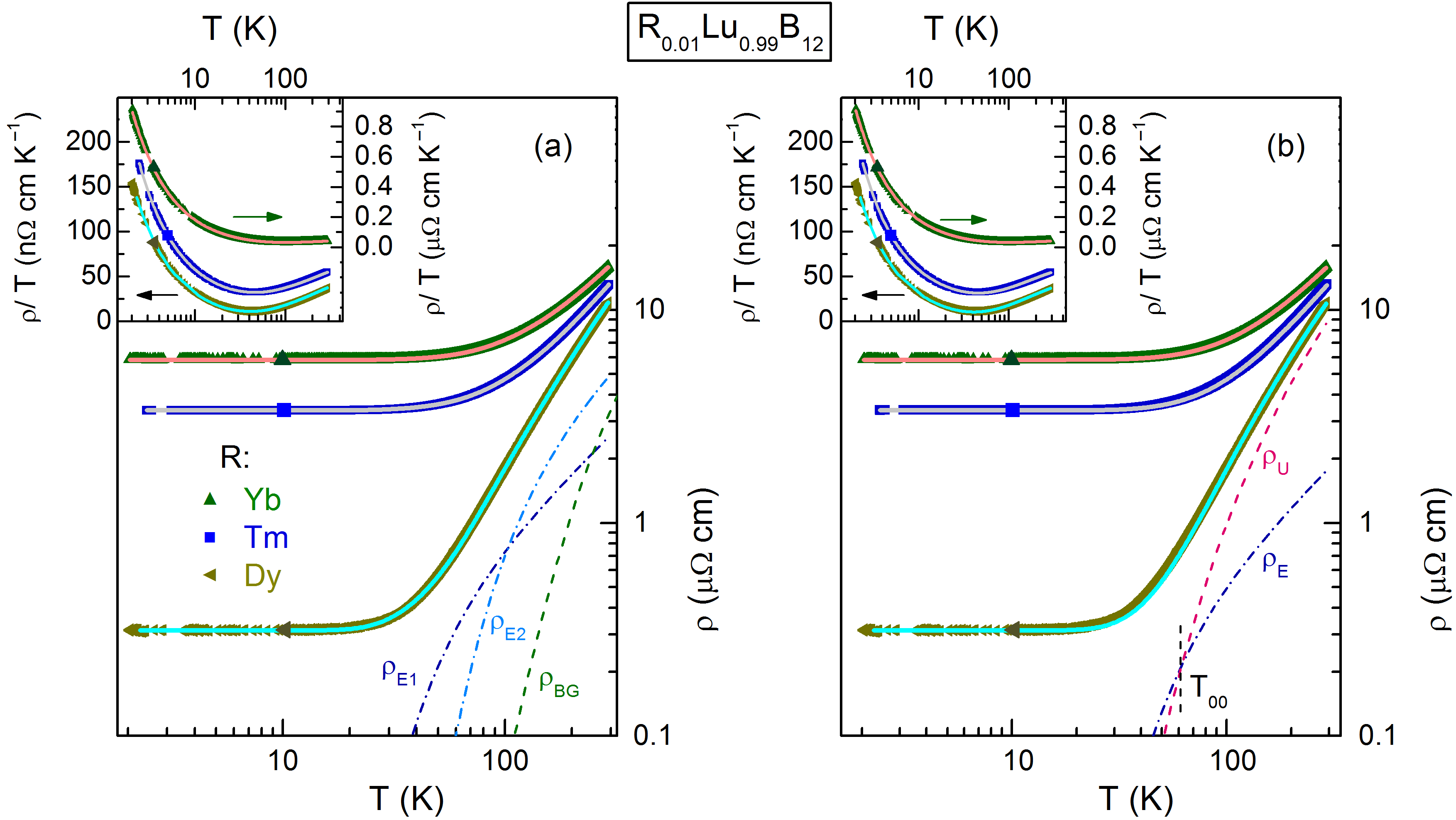}
  \caption{(Colour on-line). The fitting of $\rho$($T$) scans (solid lines) of some $R$$_{0.01}$Lu$_{0.99}$B$_{12}$ compounds ($R$-Dy, Tm, Yb), by using (a) modified method of \cite{27},  [Eqs.(\hyperref[Eq.1]{1})-(\hyperref[Eq.3]{3}), model I] and (b) model II, [Eqs.(\hyperref[Eq.1]{1}),(\hyperref[Eq.2]{2}), and (\hyperref[Eq.4]{4})]. The parameters of approximation are collected in Tables \hyperref[Tab1]{1} and \hyperref[Tab2]{2}, respectively. Separately for Dy$_{0.01}$Lu$_{0.99}$B$_{12}$ the decomposition is presented [both Bloch-Gr\"{u}neisen ($\rho_{\textmd{BG}}$) contribution in panel (a) and the term caused by Umklapp processes ($\rho_{\textmd{U}}$) in panel (b) are designated by dashed lines while Einstein modes ($\rho_{\textmd{E}}$) are displayed by dash-dotted lines, respectively]. The insets show the same fitting in coordinates $\rho$/$T$ vs $T$. The curves on each panel and also in the insets are shifted along vertical axis for convenience [except Dy$_{0.01}$Lu$_{0.99}$B$_{12}$ and Yb$_{0.01}$Lu$_{0.99}$B$_{12}$ (only in the insets)].}\label{FigX8}
   \end{center}
\end{figure}

To analyze zero-field resistivity we used the algorithm similar to that employed earlier for LaB$_6$ and LuB$_{12}$ reference materials in \cite{27, 25}. In the framework of the Matthiessen's rule the $\rho$($T$) curve is described by the sum

\begin{equation}\label{Eq.1}
\rho(T) = \rho_0 + \rho_{\textmd{BG}}(T) + \rho_\textmd{E}(T),
\end{equation}

\noindent
where $\rho$$_0$ is temperature independent residual resistivity \cite{32} caused by the scattering of electron on nonmagnetic impurities and couple components describing normal processes of electron-phonon ($e$-$ph$) scattering. As it was mentioned above the structure of $R$B$_{12}$ family may be considered as two independent sub-lattice \cite{27} giving $\rho$$_{\textmd{BG}}$($T$) Bloch-Gr$\ddot{u}$neisen term originated from rigid covalent framework of boron atoms and $\rho$$_\textmd{E}$($T$) Einstein component, caused by vibrations of RE ions ($\delta$-function in density of states). The first one is defined by the relation

\begin{equation}\label{Eq.2}
\rho_{\textmd{BG}} = (4\pi)^2 \frac{\lambda_{\textmd{tr}}\omega_\textmd{D}}{\omega_\textmd{p}^2}\left(\frac{2T}{\Theta_\textmd{D}}\right)^5
\int_0^{\Theta_\textmd{D}/2T}\frac{x^5 dx}{sh^2(x)},
\end{equation}

\noindent
where $\Theta_\textmd{D}$ is Debye temperature, $\omega_\textmd{p}$$-$ Drude plasma frequency and $\lambda_{\textmd{tr}}$$-$ electron-phonon coupling constant \cite{27}. And the second one takes a view

\begin{equation}\label{Eq.3}
\rho_{\textmd{E}} = \frac{kN}{mT\left(e^{\Theta_\textmd{E}/T}-1\right)\left(1-e^{-\Theta_\textmd{E}/T}\right)},
\end{equation}

\noindent
where $\Theta_\textmd{E}$ is Einstein temperature, $m$ $-$ atomic mass, $N$ $-$ the number of oscillators per unit volume and $k_i$ is a constant \cite{33, 34}. In order to minimize the number of variables we fixed the values of characteristic temperatures while the factors $kN$/$m$ and $\lambda_{tr}\omega_\textmd{D}$/$\omega_\textmd{p}^2$ were only changed. However, main deficiency of the model \cite{27} is that it does not always provide a sufficient convergence to experimental data with only one low-frequency Einstein mode. We digitized zero-field resistivity data for nonmagnetic ZrB$_{12}$ and UB$_{12}$ from \cite{35} and \cite{36}, respectively. It turns out, that model \cite{27} describes well the $\rho$($T$) curve of ZrB$_{12}$ with only one Einstein mode $\Theta_\textmd{E}$$^{\rho(T)}$(ZrB$_{12}$) $\approx$ 195~K, which coincides very well with the estimations from specific heat analysis $\Theta_\textmd{E1}$$^{\textmd{C(T)}}$(ZrB$_{12}$) $\approx$ 199 $-$ 202~K \cite{37}, (Table S1 in SM). On the other hand, the fitting of resistivity of UB$_{12}$ requires two Einsteinian components (compare the parameters between two models in Table S1 in SM). Such result was also reported for host LaB$_6$ and for $R$$_{0.01}$La$_{0.99}$B$_6$ family in \cite{25}. According to \cite{25} the increasing of residual resistivity in $R$$_{0.01}$La$_{0.99}$B$_6$ set leads to significant spread of the values of the factors in the framework of model \cite{27}, which is hard to expect for the systems with 1~$\%$ of doping. The usage of another one high Einstein mode improves the quality of fit and stabilizes the values of coefficients \cite{25}. This may also be illustrated for LuB$_{12}$ and LaB$_6$ reference compounds (compare Figs.S9a, S10a with Figs.S9b, S10b in SM). For example, under the condition of Einstein temperature to be the same as from specific heat analysis of Lu$^{\textmd{nat}}$B$_{12}$ (La$^{\textmd{nat}}$B$_6$) $\Theta_\textmd{E}$$^{\textmd{C(T)}}$ $\approx$ 162~K ($\Theta_\textmd{E}$$^{\textmd{C(T)}}$ $\approx$ 152.5~K), respectively \cite{6, 38} the procedure cannot describe well experimental data. If Einstein temperature is allowed to vary a better convergence may be obtained for a higher values $\Theta_\textmd{E}$$^{\rho(T)}$(LuB$_{12}$) $\approx$ 200 $-$ 202~K [$\Theta_\textmd{E}$$^{\rho(T)}$(LaB$_6$) $\approx$ 175~K], Figs.S9a, S10a (Tables S2, S3) in SM. Note that such considerable difference $\Delta\Theta_\textmd{E}$ = 38 $-$ 40~K is comparable to the effect of renormalization of $\Theta_\textmd{E}$ in dodecaborides from HoB$_{12}$ to LuB$_{12}$ \cite{26}. (In the set of RE hexaborides the Einstein temperature decreases from maximal one $\Theta_\textmd{E}$(LaB$_6$) $\approx$ 140 $-$ 152~K \cite{20, 27, 38, 39} down to $\Theta_\textmd{E}$(YbB$_6$) $\approx$ 92~K and $\Theta_\textmd{E}$(GdB$_6$) $\approx$ 91~K \cite{20, 39, 40} with $\Delta\Theta_\textmd{E}$ = 49 $-$ 61~K). Noticeable interval of deviation of temperatures in Mandrus model ($\Delta\Theta_\textmd{E}$ = 52~K, $\Delta\Theta_\textmd{D}$ = 300~K) was also reported for CaB$_4$ in \cite{41}.
Therefore an additional Einstein component $\Theta_\textmd{E2}$ $\sim$ 2$\cdot$$\Theta_\textmd{E1}$ was proposed for resistivity analysis in both LaB$_6$ and LuB$_{12}$ \cite{25}, Figs.S9b, S10b in SM. (Similar recurrent formula of Einstein frequency $\omega$$_\textmd{E}$$^{\textmd{k+1}}$ = 1.75~$\omega$$_\textmd{E}$$^{\textmd{k}}$ was derived previously for lutetium dodecaboride in \cite{8}, see the text below). The results of fitting by Eqs.(\hyperref[Eq.1]{1})-(\hyperref[Eq.3]{3}) [model I] are presented for $R$$_{0.01}$Lu$_{0.99}$B$_{12}$ by solid lines with $\Theta$$_{\textmd{E1}}$ $\approx$ 150 $-$ 156~K ($\omega$$_\textmd{E1}$ $\approx$ 12.9 $-$ 13.4~meV), $\Theta$$_{\textmd{E2}}$ $\approx$ 357 $-$ 367~K ($\omega$$_\textmd{E2}$ $\approx$ 30.8 $-$ 31.6~meV) and $\Theta$$_{\textmd{D}}$ $\approx$ 1160~K ($\omega$$_\textmd{D}$ $\approx$ 100~meV), Figs.\hyperref[FigX8]{8a}, S9b. Corresponding amplitude coefficients are collected in Table \hyperref[Tab1]{1}. Determined value of Debye temperature correlates well with the estimations obtained for $\beta$-boron phase ($\Theta$$_{\textmd{D}}$ $\approx$ 1250 $-$ 1370~K) \cite{4} as well as for other classes of rich borides $M$B$_\textmd{n}$ ($n$ = 2, 4, 6, 66) including such members as MgB$_2$ ($\Theta$$_{\textmd{D}}$$^{\rho(T)}$ = 1050~K) \cite{42}, CaB$_4$ ($\Theta$$_{\textmd{D}}$$^{\rho(T)}$ = 1230~K) \cite{41}, LaB$_6$ ($\Theta$$_{\textmd{D}}$$^{\rho(T), C(T)}$= 1160~K) \cite{27, 25, 38}, and YB$_{66}$ ($\Theta$$_{\textmd{D}}$$^{C(T)}$ = 1043~K), \cite{43}.

It should be emphasized, that additional one or even several Einstein contributions were reported previously, when analyzing heat capacity of some superconducting dodecaborides \cite{37}. On the other hand, temperature dependence of the parameter $\Theta_\textmd{D}$ described in literature \cite{34, 35, 44} as well as the effects of hybridization between the channels of scattering \cite{36} may be proposed instead a second Einstein contribution, since Matthiessen's rule often oversimplifies real processes.

Like any multi-parameter fitting the analysis of resistivity (or specific heat) in the system with complicated phonon spectrum \cite{4}, \cite{45}-\cite{47} and considerable anisotropy effects such as LuB$_{12}$ or LaB$_6$ may have several different variants of solution \cite{8, 27, 48}. All models have their advantages and deficients. In particular, a generalized Bloch-Gr$\ddot{u}$neisen equation was applied to resistivity analysis of MgB$_2$ in \cite{42}. This assumes an introducing of variable index $m$ = 3 $-$ 5 in integral expression

\begin{equation}
\left(\frac{T}{\Theta_\textmd{D}}\right)^m
\int_0^{\Theta_\textmd{D}/2T}\frac{x^m dx}{\left(e^{x}-1\right)\left(1-e^{-x}\right)},\nonumber
\end{equation}

\noindent
[instead $m$ = 5 as in Eq.(\hyperref[Eq.2]{2})]. According to \cite{42} the value $m$ = 3 gives better description of experimental data with $\Theta_\textmd{D}$$^{\rho(T)}$ = 1050~K. But the estimated value of $m$ caused some discussion about the correctness of the algorithm applied \cite{49}. The method presented for LuB$_{12}$, ZrB$_{12}$ in \cite{8} uses decomposition of total resistivity/specific heat into a sum of several Einstein components. At least 9 modes were chosen in \cite{8} to describe phonon spectra. This procedure allows to obtain high precision approximation and also to calculate the phonon density of states. At the same time it is not clear, what the minimal number of Einstein modes should be taken into account, and whether it may change with an increasing of temperature interval $i.e.$ up to 1000~K. Moreover, some low frequencies in specific heat may be erroneously interpreted as Einstenian while they may appear due to the presence of boron vacancies (defect mode), \cite{38, 50}.

Another one alternative approach of resistivity analysis of LuB$_{12}$ \cite{48} offers to consider along with the normal ($N$)

\LTcapwidth=16cm
\begin{longtable*}[!t]{cccccccc}
\caption{Analysis summary for $R$$_x$Lu$_{1-x}$B$_{12}$ materials with $x$ $\leq$  0.03 ($R$-Dy, Er, Tm, Yb and Lu) by using Eqs.(\hyperref[Eq.1]{1}), (\hyperref[Eq.3]{3}), (\hyperref[Eq.4]{4}) (with $U$-processes): $\rho_0$ is residual resistivity; ($KN$/$m$)$_1$ and $\Theta_{\textmd{E1}}$ are amplitude factor in Eq.(\hyperref[Eq.3]{3}) and Einstein temperature; A$_{\textmd{U}}$ and $T_0$ are amplitude factor and energy of substantial phonons in Eq.(\hyperref[Eq.4]{4}); $T_{00}$ is crossover point from $N$- to $U$-processes. The data for Lu$^{\textmd{nat}}$B$_{12}$ and Ho$_{0.01}$Lu$_{0.99}$B$_{12}$ crystals were taken from previous works of our group \cite{28} and \cite{31}, respectively.}\label{Tab2}\\
\hhline{========} \\
    \quad\quad \multirow{3}{*}{$R$$_x$Lu$_{1-x}$B$_{12}$}   \quad\quad &   $\rho_0$   \quad\quad &  ($KN$/$m$)$_1$   \quad\quad &    $\Theta_{\textmd{E1}}$  \quad\quad & A$_{\textmd{U}}$   \quad\quad &  $T_0$  \quad\quad &  \quad\quad & $T_{00}$   \quad\quad\\
			\\
			 \quad\quad \quad\quad &  ($\mu$$\Omega$$\cdot$cm)  \quad\quad &  (m$\Omega$$\cdot$cm$\cdot$K)  \quad\quad &    (K)  \quad\quad & (n$\Omega$$\cdot$cm/K)  \quad\quad &  (K) \quad\quad &  \quad\quad &  (K)  \quad\quad\\
			\\
\hhline{--------} \\
\quad\quad Lu$^{\textmd{nat}}$B$_{12}$  \quad\quad &  0.164  \quad\quad &  0.188  \quad\quad &    168.5  \quad\quad & 57  \quad\quad &  172 \quad\quad & \quad\quad & 69 \quad\quad \\
			\\
\quad\quad Dy$_{0.01}$Lu$_{0.99}$B$_{12}$  \quad\quad &  0.313  \quad\quad &  0.176  \quad\quad &    168.5  \quad\quad & 52.9 \quad\quad &  168.5 \quad\quad & \quad\quad & 61.7  \quad\quad \\
			\\
\quad\quad Er$_{0.01}$Lu$_{0.99}$B$_{12}$  \quad\quad &  0.417  \quad\quad &  0.149  \quad\quad &    168.5  \quad\quad & 48.5  \quad\quad &  168.5 \quad\quad & \quad\quad & 58.7 \quad\quad \\
			\\
\quad\quad Ho$_{0.01}$Lu$_{0.99}$B$_{12}$  \quad\quad &  0.722  \quad\quad &  0.164 \quad\quad &    168.5  \quad\quad & 53.7  \quad\quad &  168.5 \quad\quad & \quad\quad & 58.5  \quad\quad \\
			\\
		\quad\quad	Tm$_{0.01}$Lu$_{0.99}$B$_{12}$ \quad\quad &  0.378  \quad\quad &  0.148  \quad\quad &    168.5  \quad\quad & 49.6  \quad\quad &  168.5 \quad\quad & \quad\quad & 57.9 \quad\quad \\
			\\
			\quad\quad Yb$_{0.01}$Lu$_{0.99}$B$_{12}$  \quad\quad &  0.184  \quad\quad &  0.158 \quad\quad &    168.5  \quad\quad & 51.3 \quad\quad &  168.5 \quad\quad & \quad\quad & 58.9 \quad\quad \\
\\
\hhline{--------} \\
\quad\quad Tm$_{0.03}$Lu$_{0.97}$B$_{12}$  \quad\quad & 0.42 \quad\quad &  0.196  \quad\quad &   168.5  \quad\quad & 51.8 \quad\quad & 168.5 \quad\quad & \quad\quad & 66.9   \quad\quad \\
			\\
		\quad\quad	Yb$_{0.02}$Lu$_{0.98}$B$_{12}$ \quad\quad & 3.04 \quad\quad & 0.128 \quad\quad &  168.5  \quad\quad & 51.2 \quad\quad & 168.5 \quad\quad & \quad\quad & 52.1  \quad\quad \\
			\\
			\quad\quad Yb$_{0.03}$Lu$_{0.97}$B$_{12}$  \quad\quad &  6.28 \quad\quad  &  0.733 \quad\quad &   168.5 \quad\quad & 52.5 \quad\quad &  168.5  \quad\quad & \quad\quad & 37.8 \quad\quad \\
\\
		\hhline{========}
\end{longtable*}

\noindent
electron-phonon processes (Einstein component) additional Umklapp ($U$) ones ($\rho_\textmd{U}$). In such procedure in Eq.(\hyperref[Eq.1]{1}) Bloch-Gruneisen term is substituted by the relation

\begin{equation}\label{Eq.4}
\rho_{\textmd{U}} = A_\textmd{U} T exp\left(-\frac{T_0}{T}\right),
\end{equation}

\noindent
where $T_0$ is an energy of substantial phonons \cite{51}. Thus, the total mean free time of $e$-$ph$ scattering ($\tau_{\textmd{e-ph}}$) equals to the sum of $\tau_{\textmd{e-ph}}$$^\textmd{N}$ (diffusion motion of the electron on FS through closed area) and $\tau_{\textmd{e-ph}}$$^\textmd{U}$ (the jump with Umklapp). In Umklapp regime an electron emits or absorbs the phonons with minimal energy $T_0$. Surprisingly, but the sum of $\rho_0$, $\rho_{\textmd{E}}$, and $\rho_{\textmd{U}}$ also describes experimental data \cite{48}. Main feature of such scenario is that the value of $T_0$ $\approx$ 169~K practically coincides with Einstein parameter ($\Theta_\textmd{E}$$^{\rho\textmd{(T)}}$ $\approx$ 168.5~K) instead $\Theta_\textmd{D}$ as it is supposed to be in literature \cite{51}. According to \cite{48} a temperature of crossover $T_{00}$ (the point where $\tau_{\textmd{e-ph}}$$^\textmd{N}$ = $\tau_{\textmd{e-ph}}$$^\textmd{U}$) is only in $\approx$ 2.5 times lower than $T_0$ ($T_{00}$ $\approx$ 69~K). The ratio between $\rho_{\textmd{U}}$ and $\rho_{\textmd{E}}$ contributions achieves for Lu$^{\textmd{nat}}$B$_{12}$ the value $\rho_{\textmd{U}}$/$\rho_{\textmd{E}}$ $\sim$ $\tau_{\textmd{e-ph}}$$^\textmd{N}$/ $\tau_{\textmd{e-ph}}$$^\textmd{U}$ $\approx$ 4 at 300~K (Fig.\hyperref[FigS9]{S9c} in SM). This means, that the probability of the jump (1/$\tau_{\textmd{e-ph}}$$^\textmd{U}$) in 4 times exceeds the probability of the normal scattering (1/$\tau_{\textmd{e-ph}}$$^\textmd{N}$). In such model the spectrum of phonons is cut off by minimal frequency $T_0$ = $\Theta_{\textmd{E}}$ ($i.e.$ $\delta$-function in density of states), which turns out to be the only one energetic characteristic describing both $N$- and $U$- channels of $e$-$ph$ scattering. Moreover, the role of boron lattice is also not clear in this procedure.

We applied Eqs.(\hyperref[Eq.1]{1}), (\hyperref[Eq.3]{3}), (\hyperref[Eq.4]{4}) or model II to the analysis of resistivity for $R$$_x$Lu$_{1-x}$B$_{12}$ ($R$-Lu, Dy, Tm, Yb) compositions (Fig.\hyperref[FigX8]{8b}) and also for reference material La$^{\textmd{nat}}$B$_6$ (see Fig.S10c in SM). Under the condition of fixed Einstein temperature $\Theta_\textmd{E}$$^{\rho\textmd{(T)}}$  = $T_0$ $\approx$ 168.5~K the value of $T_{00}$ changes in the range 37.8 $-$ 66.9~K for the case of $R$$_x$Lu$_{1-x}$B$_{12}$ (Fig.\hyperref[FigX8]{8b} and Table \hyperref[Tab2]{2}). In Yb$_{x}$Lu$_{1-x}$B$_{12}$ series $T_{00}$ decreases from 58.9~K down to 37.8~K (Table \hyperref[Tab2]{2}). It should be emphasized, that the parameter $T_{00}$ is very sensitive to the position of residual resistivity $\rho_0$. For instance, the substitution of $\rho_0$ from 6.28~$\mu$$\Omega$$\cdot$cm (the position of the minimum) to an arbitrary chosen 6.18~$\mu$$\Omega$$\cdot$cm in Yb$_{0.03}$Lu$_{0.97}$B$_{12}$ is accompanied by the change of $T_{00}$ by more than 10~K ($T_{00}$ $\approx$ 56.2~K). If both parameters $\Theta_\textmd{E}$ and $T_0$ are allowed to vary a considerable discrepancy of the values is obtained ($\Theta_\textmd{E}$ $\approx$ 190 $-$ 222~K, $T_0$ $\approx$ 245 $-$ 265~K and $T_{00}$ $\approx$ 171 $-$ 218~K).
In the case of LaB$_6$ the fit presented in Fig.S10c [$T_0$ = $\Theta_\textmd{E}$(LaB$_6$) $\approx$ 155~K, Table S3 in SM] does not give sufficient convergence to experimental data especially in the range 38~K $<$ $T$ $<$ 115~K. According to Fig.S10c the problem of second Einstein component is not solved for LaB$_6$, even when $U$-processes are taken into account \cite{48}. Besides, the temperature of crossover $T_{00}$(LaB$_6$) $\approx$ 189~K (Table S3 in SM) is almost in 2 times higher in comparison with the point $T^*$ $\approx$ 92~K \cite{38}, which may be attributed to the temperature of cage glass state formation (see the discussion in Sect.\hyperref[Sec4p4]{4.4}).

Thus, the analysis of resistivity/specific heat is the attempt to describe complicated phonon spectrum. In our opinion, from all above methods the modified model of \cite{27} turns out to be the most preferable, because it satisfies criteria of simplicity (describes the $e$-$ph$ scattering without cutting of phonon spectrum) and universality. For example, the fit may be extrapolated to higher temperatures up to 1000~K. In addition, it may be applied to the analysis of resistivity for other members of the class of rich borides  even in spite of the presence $U$-processes in the system.

Another feature to be discussed is the effect, when zero-field resistivity of La$^{\textmd{nat}}$B$_6$ exceeds analogous data of LuB$_{12}$ only in the range $\approx$ 50 $-$ 75~K. The details are presented in \cite{2, 25}. Note that the difference in Einstein temperatures between Lu$^{\textmd{nat}}$B$_{12}$ and La$^{\textmd{nat}}$B$_6$ objects $\Delta$$\Theta_\textmd{E}$$^{\textmd{C(T)}}$ $\approx$ 9.8~K  is rather small \cite{6, 38}. It is less than the difference in $\Theta_\textmd{E}$ values between corresponding reference system LuB$_{12}$/LaB$_6$ and the nearest neighbor in the set (YbB$_{12}$/CeB$_6$): $\Delta$$\Theta_\textmd{E}$(YbB$_{12}$ $-$ LuB$_{12}$) = 13 versus $\Delta$$\Theta_\textmd{E}$(LaB$_6$ $-$ CeB$_6$) = 14, respectively \cite{20, 26}. However, the shape of $\rho$($T$) curves is different. Model I with two Einstein components explains such effect as redistribution between the amplitude coefficients ($kN$/$m$)$_i$ in Eq.(\hyperref[Eq.2]{2}), see \cite{25}.

\subsection*{\emph{4.2. Transport anisotropy and the effects of Fermi surface topology.}}\label{Sec4p2}

Significant anisotropy in magnetotransport has been reported for both LuB$_{12}$ and LaB$_6$ reference systems in \cite{3, 28, 52}. For instance, the ratio of MR$_+$ between \textbf{H}$\|$[100] and \textbf{H}$\|$[111] or \textbf{H}$\|$[110] field directions achieves for Lu$^{\textmd{nat}}$B$_{12}$ the values 5/3.7 at the point $T$ = 4.2~K and $H$ = 80~kOe (A$_{\textmd{MR}}$$^{\textmd{max}}$ $\approx$ 480~$\%$, the inset in Fig.\hyperref[FigX6]{6b}), \cite{28}. Moreover, unusually strong difference of Hall mobility (by a factor of about 6) was also registered in the same crystal between the orientations \textbf{H}$\|$[100] and \textbf{H}$\|$[110], \cite{28}. (In comparison, the data digitized for LaB$_6$ from \cite{52} give $A_{\textmd{MR}}^{\textmd{max}}$ $\approx$ 480~$\%$ for  current geometry \textbf{I}$\|$[110] at the point $T$ = 1.4~K and $H$ = 56.7~kOe). The doping of LuB$_{12}$ with 1 $-$ 3~$\%$ of RE impurity strongly affects on the shape of $\Delta\rho/\rho$ = $f$($\varphi$) curves resulting also to the depression of the amplitude of MR$_+$ and its anisotropy (Fig.\hyperref[FigX6]{6}). Secondly, the shape of Fermi surface of both LaB$_6$ and LuB$_{12}$ reference systems also suggests the presence of open orbits \cite{3, 52, 53}. In particular, FS of LaB$_6$ consists of the ellipsoids centered at the X-points and connected by short necks \cite{52}-\cite{FS2}. Alternative model of FS of LaB$_6$ based mainly on the results of acoustic de Haas-van Alphen effect (dHvA) studies suggests also the presence of small ellipsoid-electron-pockets which are overlap on the necks \cite{FS3}-\cite{FS5}.  Being more complicated FS of LuB$_{12}$ consists of two conduction bands. The upper band (electronic sheet) corresponds to a simply ($\textmd{'}$pancake$\textmd{'}$-like) connected Fermi surfaces \cite{1, 3, 54}, while the lower one (hole sheet) shows a multiple connected $\textmd{'}$monster$\textmd{'}$ shape similar to noble metals \cite{3}. Open orbits occur for hole sheet of FS \cite{3}. Note that the saturation of magnetoresistance was reported for LuB$_{12}$ only in configuration \textbf{I}$\|$[110] and \textbf{H}$\|$[111], whereas for the rest main directions of magnetic field (\textbf{H}$\|$[100] and \textbf{H}$\|$[110]) unsaturated behavior was observed \cite{3}. However, in \cite{28} the tendency to saturation of MR was also reported for geometry \textbf{I}$\|$[110] and \textbf{H}$\|$[110]. Here, for Dy$_{0.01}$Lu$_{0.99}$B$_{12}$ composition we obtained the results similar to \cite{3} with almost quadratic asymptotic $\sim$ $H^2$ of positive magnetoresistance in both directions \textbf{H}$\|$[100] and \textbf{H}$\|$[110], Fig.\hyperref[FigX7]{7a}. In the framework of classical physics of metals \cite{55} a strong increase of MR in external magnetic field without saturation should indicate the presence of open orbits.

As one of the arguments for using the conception of additional channel of scattering (dynamic charge stripes) it was pointed out in \cite{28}, that high-field condition is not valid. Indeed, the number of cyclotron motions between scattering events ($i.e.$ the parameter $\omega_\textmd{c}$$\tau$) should satisfy the inequality $\omega_\textmd{c}$$\tau$ $\gg$ 1 or $r_\textmd{L}$ $\ll$ l (where $\omega_\textmd{c}$ = $eH$/$m^*c$ is cyclotron frequency, $\tau$ $-$ scattering lifetime, as well as $r_\textmd{L}$ $-$ Larmor radius and $l$ $-$ mean free path). The data of \cite{28} showed, that this condition was not true in the set of LuB$_{12}$ single crystals below 120~kOe even for Lu$^{\textmd{nat}}$B$_{12}$ the purest among of them. The conception \cite{28} explains unsaturated $\sim$ $H^2$ growth of positive MR along \textbf{H}$\|$[110] direction by transition from closed to quasi-open orbit, for which the mean free path equals to $\sim$ 5 $-$ 10~$a$. According to the theorem of Lifshits-Peschanski such trajectory may remain closed despite its length.

On the other hand Hall effect measurements \cite{28} give only an averaged value of the parameter $\omega_\textmd{c}$$\tau$ and don't take into account the difference between charge carriers with light and heavy cyclotron masses. According to dHvA experiments the complex spectra of the frequencies is realized for RE dodecaborides including LuB$_{12}$, for which the effective cyclotron masses of the electrons change in the range $\approx$ 0.38 $-$ 2.23~$m_0$ \cite{3}. (For example, in metallic UB$_{12}$ the high field condition is satisfied along the direction $\langle$100$\rangle$ at the point 70~kOe only for the carriers with light cyclotron masses \cite{56}). According to alternative estimations in \cite{28} there is at least one group of carriers in LuB$_{12}$ ('four cornered rosette' R$_{100}$ orbit, with effective mass $m^*$$_{\alpha2}$ $\approx$ 0.53~$m_0$ and Hall mobility $\mu_{\alpha2}$ $\approx$ 7100~cm$^2$V$^{-1}$s$^{-1}$), for which the number of cyclotron motions $\omega_\textmd{c}$$\tau$ achieves a noticeable value of $\sim$ 2$\pi$ at 90~kOe. Thus, the appealing to non-fulfillment of high-field condition in LuB$_{12}$/$R$$_x$Lu$_{1-x}$B$_{12}$ materials requires additional verification.

\begin{figure}[htpb]
\begin{center}
\includegraphics[width = 14 cm]{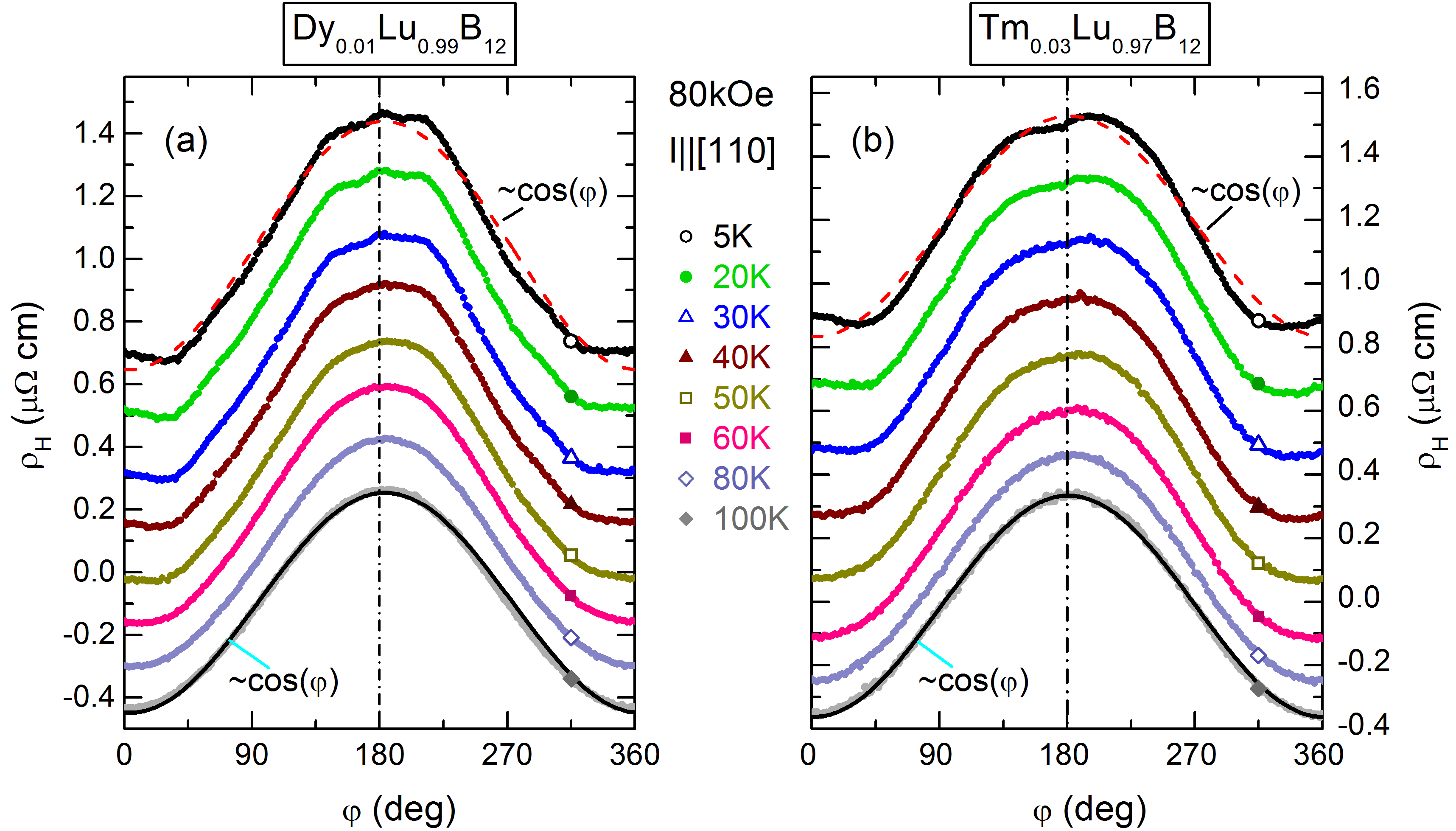}
  \caption{(Colour on-line). Angular dependencies of Hall resistivity $\rho_\textmd{H}$($\varphi$) of (a) Dy$_{0.01}$Lu$_{0.99}$B$_{12}$ and (b) Tm$_{0.03}$Lu$_{0.97}$B$_{12}$ compositions, measured at constant magnetic field $H$ = 80~kOe (current geometry \textbf{I}$\|$[110]) for the isotherms from the range 5 $-$ 100~K. The curves on each panel are shifted along vertical axis for convenience. Solid and dashed lines display the fit by harmonic law $\rho_\textmd{H}$($\varphi$) $\sim$ cos($\varphi$), see Sect.\hyperref[Sec4p2]{4.2} for details.}\label{FigX9}
   \end{center}
\end{figure}

Another evidence of the presence of an additional scattering channel caused by the formation of dynamic charge stripes is the deviation of the experimental data of the Hall resistivity $\rho_\textmd{H}$($\varphi$) from the harmonic law $\rho_\textmd{H}$($\varphi$) $\sim$ cos($\varphi$) below $\sim$ 100 $-$ 150~K \cite{28}. We also measured Hall resistivity simultaneously with magnetoresistance in the same current geometry \textbf{I}$\|$[110]. Figure \hyperref[FigX9]{9} shows $\rho_\textmd{H}$($\varphi$) dependencies obtained for (\hyperref[FigX9]{a}) Dy$_{0.01}$Lu$_{0.99}$B$_{12}$ and (\hyperref[FigX9]{b}) Tm$_{0.03}$Lu$_{0.97}$B$_{12}$ compositions at constant magnetic field $H$ = 80~kOe. Our data allow observing, that Hall resistivity is described by asymptotic $\sim$ cos($\varphi$) above the point $\sim$ 60~K (solid lines in Fig.\hyperref[FigX9]{9}), while below this temperature there is a deviation from harmonic law (dashed lines in Fig.\hyperref[FigX9]{9}) with a change of shape of $\rho_\textmd{H}$($\varphi$) curves. Consequently, additional features are registered near \textbf{H}$\|$[100] direction at low $T$-limit. Such behavior correlates well with the results of both Fig.\hyperref[FigX5]{5a} and Sect.\hyperref[Sec4p3]{4.3}. Thus, the anomalies appear on the angular dependencies of two components of resistivity tensor (transverse magnetoresistance and Hall resistivity) at similar narrow range of angles at the same constant magnetic field (Figs.\hyperref[FigX5]{5a}, \hyperref[FigX6]{6a}, \hyperref[FigX7]{7b}, \hyperref[FigX9]{9}), that is expected for the effects of FS topology \cite{51, 55}. In the case of Lu$^{\textmd{nat}}$B$_{12}$ host system there are several additional peculiarities on MR$_+$ dependencies. In particular, except beak-shaped singularities observed for \textbf{H}$\|$[110] two more features (narrow dips along \textbf{H}$\|$[100] and small twin-anomalies at $\varphi$ $\approx$ 76 $-$ 81$^{\circ}$ $\pm$ $\Delta\varphi$ and $\varphi$ $\approx$ 100 $-$ 105$^{\circ}$ $\pm$  $\Delta\varphi$) are visible on the curves $\Delta\rho/\rho$ = $f$($\varphi$) at $H$ = 80~kOe \cite{28}. One of them (narrow dips along \textbf{H}$\|$[100]) corresponds to step-like peculiarities on angular dependencies of Hall resistivity \cite{28}. It will be shown below, that conception of dynamic charge stripes may be excluded, when explaining observed phenomena.

\subsection*{\emph{4.3. $H$ $-$ dependencies of magnetoresistance. Kohler's rule and drift mobility.}}\label{Sec4p3}

Let us compare field dependencies of magnetoresistance measured for $R$$_x$Lu$_{1-x}$B$_{12}$ family (Figs.\hyperref[FigX4]{4}, \hyperref[FigX7]{7c}$-$\hyperref[FigX7]{7d} and Figs.S5$-$S8 in SM) with the similar data published recently for LaB$_6$-based hexaborides with the formula $R$$_{0.01}$La$_{0.99}$B$_6$ ($R$-La, Ce, Pr, Nd, Eu, Gd, and Ho) in \cite{25}. One of the results detected for the latter class of materials is significant influence of magnetic impurity on MR data. In particular, several different types of MR behavior were observed in \cite{25} including a dominance of positive linear asymptotic ($\Delta\rho$/$\rho$ $\sim$ $AH$) in a wide interval of fields $H$ $\leq$ 70~kOe (Eu$_{0.01}$La$_{0.99}$B$_6$ and Nd$_{0.01}$La$_{0.99}$B$_6$), and crossover from positive $T$ $>$ $T_\textmd{{inv}}$ to negative ($T$ $<$ $T_\textmd{{inv}}$) regime (Ce$_{0.01}$La$_{0.99}$B$_6$ and Ho$_{0.01}$La$_{0.99}$B$_6$). However, the changing of RE impurity from $R$-Dy to Yb in $R$$_x$Lu$_{1-x}$B$_{12}$ set does not produce significant effect on the asymptotic of magnetoresistance (only on the amplitude) being described by dominant positive quadratic component (above $\approx$ 40 $-$ 50~kOe) for all the compounds under investigation.

\begin{figure}[htpb]
\begin{center}
\includegraphics[width = 14cm]{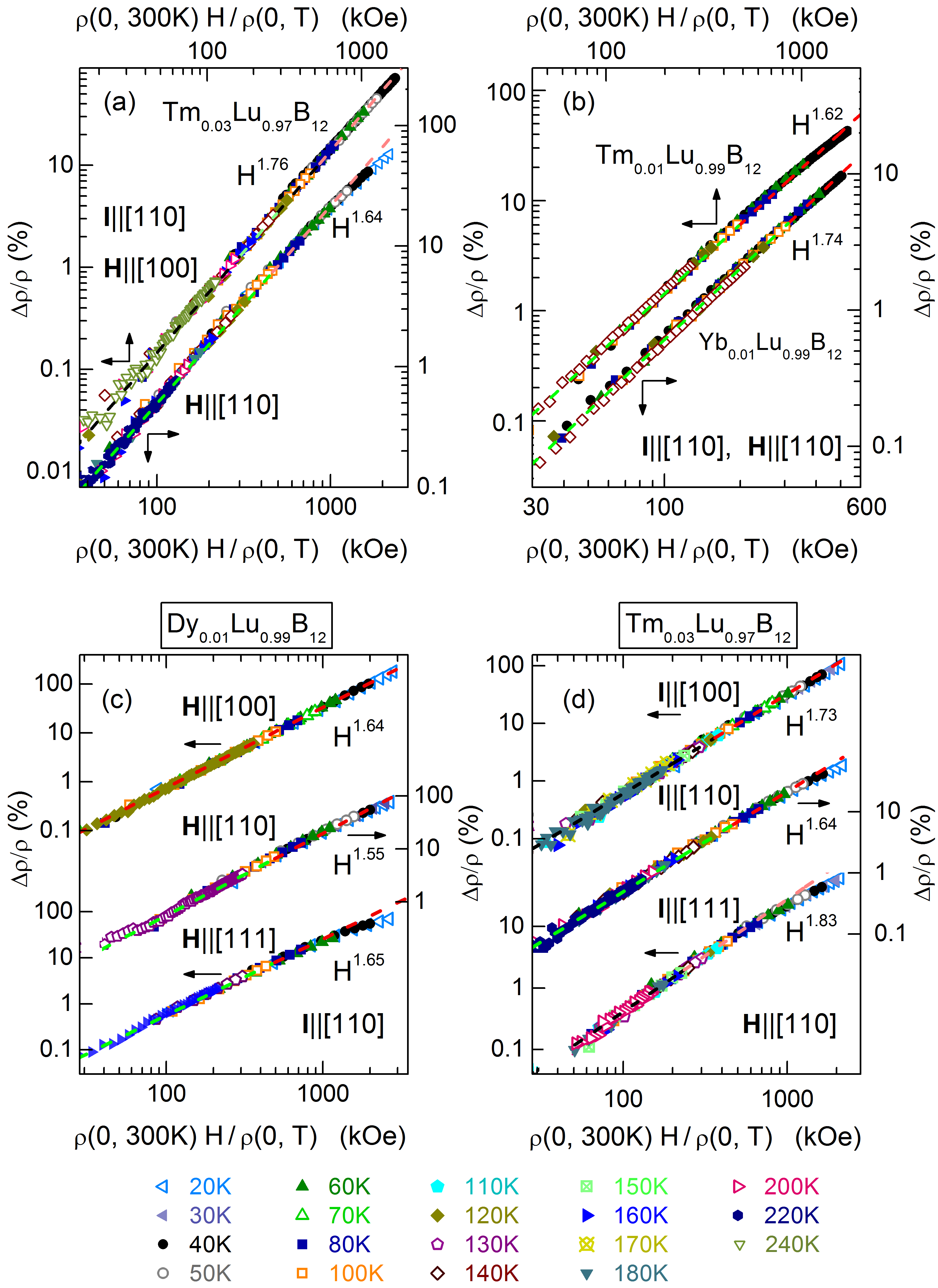}
 \caption{(Colour on-line). Kohler's plot of magnetoresistance $\Delta\rho$($H$, $T$)/$\rho$ = $f$[$\rho$(0, 300~K)$H$/$\rho$(0, $T$)] for
(a),(d) Tm$_{0.03}$Lu$_{0.97}$B$_{12}$,  (b) Tm$_{0.01}$Lu$_{0.99}$B$_{12}$, Yb$_{0.01}$Lu$_{0.99}$B$_{12}$, and (c) Dy$_{0.01}$Lu$_{0.99}$B$_{12}$ compositions. Separately, panels (a), (c)-(d) show data measured in various experimental geometries of (a), (c) field with fixed current orientation \textbf{I}$\|$[110] and of (d) current with fixed field orientation \textbf{H}$\|$[110]. Dashed lines on each panel represent the approximation by power law $\Delta\rho$/$\rho$ $\sim$ $H^a$ with the index $a$ $\approx$ 1.55 $-$ 1.83 (see the text for details). }\label{FigX10}
 \end{center}
\end{figure}

This fact allows proposing scaling behavior within the framework of semi-empirical Kohler's rule (KR). Here, the simplest form of KR is used \cite{51}. Based on the idea, that there is a single momentum-independent mean free path ($l$) for all electrons, it is expressed by the relation

\begin{figure*}[!b]
\hspace{-1cm}\includegraphics[width = 18cm]{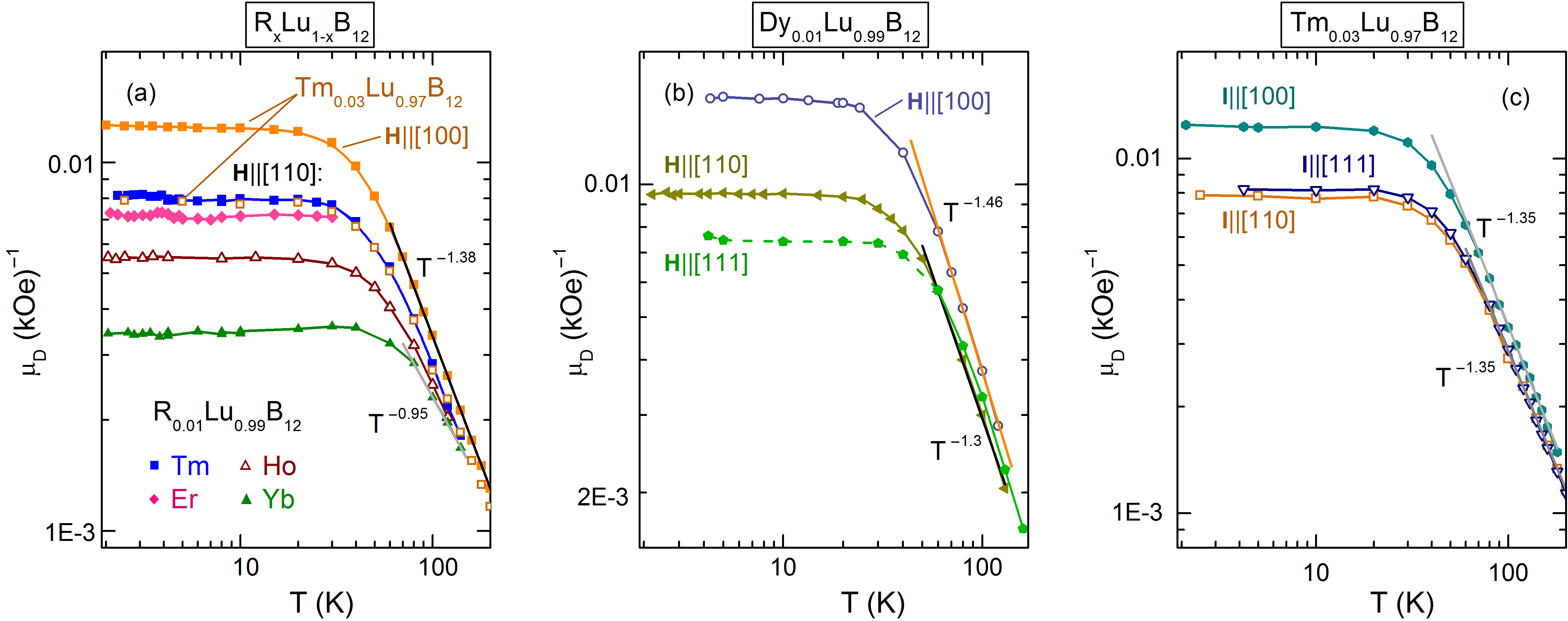}
   \parbox{18cm}{\caption{(Colour on-line). Reduced drift mobility of the charge carriers $\mu_\textmd{D}$ vs temperature for $R$$_x$Lu$_{1-x}$B$_{12}$ family. Solid lines represent the approximation by power law $\mu_\textmd{D}$($T$) $\sim$ $T^{-\alpha}$ with the index $\alpha$ $\approx$ 0.95 $-$ 1.46. Separately for compositions Dy$_{0.01}$Lu$_{0.99}$B$_{12}$ and Tm$_{0.03}$Lu$_{0.97}$B$_{12}$  the curves $\mu_\textmd{D}$($T$) were evaluated for different (a)-(b) field (\textbf{H}$\|$[100], \textbf{H}$\|$[110], \textbf{H}$\|$[111]) and (c) current (\textbf{I}$\|$[100], \textbf{I}$\|$[110], \textbf{I}$\|$[111]) orientations.}}\label{FigX11}
\end{figure*}

\begin{equation}\label{Eq.5}
\Delta\rho (0, T)/\rho = f\left(\frac{\rho(0, 300~\textmd{K})}{\rho(0, T)}H\right) \propto \left(\frac{l}{r_\textmd{L}}\right)^2,
\end{equation}

\noindent
where $r_\textmd{L}$ is the Larmour radius of electron precession in the magnetic field and $f$ $-$ the function depending on the nature of the metal. In Fig.\hyperref[FigX10]{10} we plotted magnetoresistance of  (\hyperref[FigX10]{a}), (\hyperref[FigX10]{d}) Tm$_{0.03}$Lu$_{0.97}$B$_{12}$, (\hyperref[FigX10]{b}) Tm$_{0.01}$Lu$_{0.99}$B$_{12}$,Yb$_{0.01}$Lu$_{0.99}$B$_{12}$, and (\hyperref[FigX10]{c})  Dy$_{0.01}$Lu$_{0.99}$B$_{12}$ by using Eq.(\hyperref[Eq.5]{5}) representation.
Separately, panels (\hyperref[FigX10]{a}), (\hyperref[FigX10]{c})-(\hyperref[FigX10]{d}) show the data measured in various experimental geometries of (\hyperref[FigX10]{a}), (\hyperref[FigX10]{c}) magnetic field with the same current direction (\textbf{I}$\|$[110]) and of (\hyperref[FigX10]{d}) current with the same orientation of magnetic field (\textbf{H}$\|$[110]).  [Similar analysis is applied to Ho$_{0.01}$Lu$_{0.99}$B$_{12}$ system in Fig.S11]. In double logarithmic plot $\Delta\rho$/$\rho$ data collapse into one curve $\sim$ $H^a$ with the index $a$ $\approx$ 1.55 $-$ 1.83 (see dashed lines in Fig.\hyperref[FigX10]{10}). Good scaling behavior of MR$_+$ is observed in a wide interval of temperatures 40 $-$ 240~K [60 $-$ 160~K for Dy$_{0.01}$Lu$_{0.99}$B$_{12}$ in geometry \textbf{H}$\|$[111] (Fig.\hyperref[FigX10]{10c}), 80 $-$ 200~K for Tm$_{0.03}$Lu$_{0.97}$B$_{12}$ in geometry \textbf{I}$\|$[111] (Fig.\hyperref[FigX10]{10d})] and 40 $-$ 120~K for Ho$_{0.01}$Lu$_{0.99}$B$_{12}$ (Fig.S11 in SM) corresponding to the dominance of electron scattering on Einstein phonons (Sect.\hyperref[Sec4p1]{4.1}). It should be emphasized, that this result is received among the other things in the $T$-range corresponding to possible charge stripes formation $T$ $<$ $\sim$ 150~K (Figs.\hyperref[FigX10]{10c}, \hyperref[FigX10]{10d}). But we don't see any deviation from Kohler's rule of $R$$_{x}$Lu$_{1-x}$B$_{12}$ in above interval for all directions of vectors of magnetic field or current. Moreover, taking into account the discussion about probable dominance of $e$-$ph$ Umklapp processes at $T$ $>$ $T_{00}$ $\approx$ 37.8 $-$ 69~K (Sect.\hyperref[Sec4p1]{4.1}) this result  seems for us to be another one argument in favor of using modified model \cite{27} instead model II for resistivity approximation. It is very likely, that KR is also fulfilled for LuB$_{12}$ host material. We digitized MR data for nonmagnetic metal UB$_{12}$ from \cite{36} and obtain that they also collapse into one curve $\sim$ $H^{1.6}$ above 40~K.

However, below the point $\approx$ 25 $-$ 30~K (60~K for \textbf{H}$\|$[111] in Dy$_{0.01}$Lu$_{0.99}$B$_{12}$ or $\sim$ 80~K for \textbf{I}$\|$[111] in Tm$_{0.03}$Lu$_{0.97}$B$_{12}$) there is a deviation from Kohler's asymptotic since the scattering of electrons by both magnetic and nonmagnetic impurities is a chief component to zero-field resistivity of investigated $R$$_{x}$Lu$_{1-x}$B$_{12}$. It is worth noting, that in above interval the MR curves $\Delta\rho$/$\rho$ = $f$($H$) practically coincide with each other except the case of Yb$_{x}$Lu$_{1-x}$B$_{12}$ (see Fig.S7 in SM). Taking into account the results of zero-field resistivity analysis in Sect.\hyperref[Sec4p1]{4.1} it may be assumed that KR is also satisfied for other current and field geometries in LuB$_{12}$.

The results above indicate the absence of additional channel of scattering caused by the formation of dynamic charge stripes below 150~K \cite{28}. Thus, it may be proposed, that MR anisotropy is originated in $R$$_{x}$Lu$_{1-x}$B$_{12}$ due to the anisotropy of $e$-$ph$ scattering on the one hand and the effects of Fermi surface topology on the other hand (Sect.\hyperref[Sec4p2]{4.2}). The latter become noticeable, when the amplitude of Einsteinian contributions Eq.(\hyperref[Eq.3]{3}) decreases with cooling, $i.e.$ below 40 $-$ 60~K/80~K (depending on experimental geometry), which is confirmed in experiment (see Figs.\hyperref[FigX5]{5a}, \hyperref[FigX8]{8a} and \hyperref[FigX9]{9}).

Summarizing the analysis of positive quadratic component of magnetoresistance we estimated reduced drift mobility of the charge carries ($\mu_\textmd{D}$) as a function of temperature for $R$$_{0.01}$Lu$_{0.99}$B$_{12}$ ($R$-Ho, Er, Tm, Yb), Tm$_{0.03}$Lu$_{0.97}$B$_{12}$ compounds (Fig.\hyperref[FigX11]{11a})
and separately for different geometries of magnetic field (Fig.\hyperref[FigX11]{11a, 11b}) in
Dy$_{0.01}$Lu$_{0.99}$B$_{12}$, Tm$_{0.03}$Lu$_{0.97}$B$_{12}$ and current (Fig.\hyperref[FigX11]{11c}) in Tm$_{0.03}$Lu$_{0.97}$B$_{12}$.
As can be seen from Fig.\hyperref[FigX11]{11} $\mu_\textmd{D}$ obeys a power law $\sim$ $T^{-\alpha}$ above 60~K with the index $\alpha$ $\approx$ 0.95 $-$ 1.46. Index $\alpha$ depends on the orientation of magnetic field  (for Dy$_{0.01}$Lu$_{0.99}$B$_{12}$  $\alpha$$^{\textbf{H} \| [110]}$ $\approx$ 1.3 $<$ $\alpha$$^{\textbf{H} \| [111]}$ $\approx$ 1.39 $<$ $\alpha$$^{\textbf{H} \| [100]}$ $\approx$ 1.46) and does not change for different current directions (Fig.\hyperref[FigX11]{11c}). The results obtained correlate well with previous investigations of magnetotransport in Ho$_{x}$Lu$_{1-x}$B$_{12}$ where a good agreement between $\mu_\textmd{D}$ and Hall mobility $\mu_\textmd{H}$ = $R_\textmd{H}$/$\rho$ has been reported \cite{31}. According to \cite{31} both characteristics described by the relation $\mu_\textmd{D}$ $\sim$ $\mu_\textmd{H}$ $\sim$ $T^{-\alpha}$ have nearly equal indexes ($\alpha_{\textmd{D}}$ and $\alpha_{\textmd{H}}$) e.g. $\alpha$(Ho$_{0.1}$Lu$_{0.9}$B$_{12}$) $\approx$ 1.6 $-$ 1.7 and $\alpha$(Ho$_{0.5}$Lu$_{0.5}$B$_{12}$) $\approx$ 1.3. It is worth noting, that Hall effect studies of Lu$^{\textmd{nat}}$B$_{12}$ crystal \cite{28} give a slightly higher value of $\alpha$ $\approx$ 1.75 for $\mu_\textmd{H}$.

\subsection*{\emph{4.4. Discussion.}}\label{Sec4p4}

It is necessary to mention, that there are several scenarios in literature describing $R$B$_{12}$ (and likely $R$B$_6$) cubic frame-clustered systems. In particular, a cage glass state (static random displacement of RE ions in B$_{24}$ truncated cuboctahedra) has been proposed for LuB$_{12}$ below $T$$^*$ $\approx$ 60~K due to $n_{\textmd{vac}}$ $\leq$ 4~$\%$ vacancies at boron positions ($T$$^*$ $\approx$ 92~K with only $n_{\textmd{vac}}$ $\approx$ 1.5~$\%$ for La$^{\textmd{nat}}$B$_{6}$). It is accompanied by probable Jahn-Teller structural instabilities of rigid boron cage \cite{6, 48} and also by the formation of the collective mode (overdamped oscillator) produced by 70~$\%$ of nonequilibrium electrons \cite{57}. In this approach the effects of considerable anisotropy in magnetotransport detected for LuB$_{12}$ were explained among other things by the formation of dynamic charge stripes (charge fluctuations) in $fcc$ lattice along the direction $\langle$110$\rangle$ (at least for trivalent RE dodecaborides) \cite{28, 48}. It was also reported, that possible emergence of dynamic charge stripes is equivalent to $ac$ current with a frequency $\approx$ 200~GHz \cite{58}. An average length of such stripe was estimated for YbB$_{12}$ as $\langle$$L$$\rangle$ $\sim$ 100~$\textmd{\AA}$ or $\approx$ 13~$a$ \cite{59}. Note that from the point of view of the authors \cite{6, 28, 48} this model may also be extrapolated to the class of RE hexaborides.

Interestingly, the formation of the stripes was discussed for the systems with layered structures such as cuprates \cite{60}, nickelates \cite{61} and also for many strongly correlated classes with complex hierarchy of interactions and rich $x$-$T$-$H$ phase diagrams \cite{62, 63}. Therefore in order to apply mentioned scenario to $R$B$_{12}$/$R$B$_6$ families in a view of the authors \cite{6, 48} it is important to search  evidences of structural instability or even symmetry lowering. As additional arguments one may consider the results of Raman scattering investigations in LaB$_6$ \cite{47}, where it was reported that lanthanum hexaboride does not possess cubic symmetry (see the text below). Moreover, density functional theory calculations of phonon dispersions of several hexaborides show, that models with lower symmetry (such as $P$4/$mmm$ superlattice) gives better convergence at least for YB$_6$ \cite{64}.

Another opportunity for anisotropy development in the $R$B$_6$ systems was proposed for CeB$_6$ \cite{65A}. In this material the phase with hidden magnetic order [antiferroquadrupolar (AFQ) phase] exists and the transition from paramagnetic to AFQ phase is accompanied by a spin fluctuation transition consisting in the onset of spin fluctuations anisotropy. Scattering on anisotropic magnetic fluctuations gives rise to the anisotropy of magnetoresistance with 180$^{\circ}$ angular symmetry, which is often referred as electron nematic effect \cite{65A}.

An alternative paradigm gives different interpretation of experimental results. For instance, magnetic anisotropy in Ce$_{1-x}$La$_x$B$_6$ was explained by using local two-site model, which does not violate cubic symmetry \cite{67}. Here, it is important to note about the problem of possible phase transition in LaB$_6$ under external pressure \cite{68, 69}. Indeed, the change from cubic to orthorhombic crystal structures was proposed at 10~GPa in \cite{68}. The verification of this result performed for polycrystalline samples on the basis of high-resolution X-ray diffraction studies and Raman spectroscopy measurements showed no evidence of structural or electronic phase transition pointing, that cubic symmetry survives in LaB$_{6}$ at least up to 25~GPa at room temperature \cite{69}. Moreover, even the assumption of a possible transition to cage glass state (CGS) in LuB$_{12}$/LaB$_6$ is the subject of debates \cite{70}-\cite{72}. According to \cite{70} a low frequency peak registered in Raman spectra of LuB$_{12}$ may be caused by intraband electron transitions. The formation of CGS in $R$B$_{12}$ was not confirmed in EXAFS analysis due to the sensitivity limits \cite{72}. Indeed, the estimations obtained from heat capacity \cite{38} and EXAFS experiments \cite{26} showed, that the fraction of displaced ions in LuB$_{12}$ and LaB$_{6}$ materials is low [$Q$(LuB$_{12}$) $\leq$ 3.6~$\%$ and $Q$(LaB$_6$) $\leq$ 6~$\%$]. The rest 94 $-$ 96~$\%$ RE ions stay in initial centrosymmetric positions. Defect mode caused by the vacancies gives only small correction to the dominant Einstein term in heat capacity analysis of both LaB$_6$ and LuB$_{12}$ systems \cite{6, 38, 50}. The appearance of significant fraction of displaced ions contradicts to the criterion of stability of $R$B$_6$/$R$B$_{12}$ structure (see Sect.\hyperref[Sec.1]{1}). Thus, the existence of CGS in rich borides remains a disputable question and further investigations are required.

As extra arguments for this scenario one may also consider the results of NMR experiments performed for YB$_6$ in \cite{73}, where the transition to static disorder was not confirmed. The interpretation given instead includes the transition to dynamic disordered state in YB$_6$ at low temperatures $T$ $<$ 50~K \cite{73}. Moreover, according to numerous X-ray/neutron scattering studies as well as various other experiments obtained for $R$B$_6$ class by different groups \cite{19, 20, 39, 69}, \cite{74}-\cite{82} the cubic symmetry is preserved in a wide range of temperatures corresponding to paramagnetic/diamagnetic state. Any lattice distortions (or even structural transitions) were reported in the vicinity of (or simultaneously with) magnetic transitions. In particular, intermediate ferroquadrupolar phase accompanied by a structural change was observed in antiferromagnets (AFM) HoB$_6$ and DyB$_6$ [$T_\textmd{N}$(HoB$_6$) = 5.7~K, $T_\textmd{N}$(DyB$_6$) = 23 $-$ 26~K] below the point $T_\textmd{Q}$(HoB$_6$) = 6.1~K, $T_\textmd{Q}$(DyB$_6$) = 31 $-$ 32~K \cite{83}-\cite{85}. X-ray diffraction studies of DyB$_6$ \cite{84} proves, that cubic phase coexists with large lattice distortions of rhombohedral type in the range $T_\textmd{N}$ $-$ $T_\textmd{Q}$, while only below Neel point the structure changes to rhombohedral one. Remarkably, that tetragonal symmetry was proposed in AFM state of TbB$_6$ \cite{86} and structural distortion associated with the incommensurate to commensurate phase transition was discussed for PrB$_6$ in \cite{87}. In $S$-system ferromagnet EuB$_6$ (the configuration of $^8$$S$$_{7/2}$), which in first order is not influenced by strong crystal-electric field effects and Jahn-Teller distortions, a very large lattice response was observed only in paramagnetic vicinity of Curie point \cite{88}. According to \cite{88} the lattice effects in EuB$_6$ originate in the magnetically driven delocalization of charge carriers (the percolating magnetic polarons). In AFM $S$-object GdB$_6$ ($T_{\textmd{N1}}$ = 15 $-$ 16~K \cite{B1, B2}), where the amplitude of thermal vibrations of Gd$^{3+}$ ions achieves large values \cite{B3}, a magnetic structure coexist with displacement waves only below Neel point \cite{B4}.

Finally, it is not clear how possible formation of dynamic charge stripes correlates with short-range magnetic effects registered in paramagnetic state for both hexa- and dodecaboride classes \cite{89}-\cite{92}. For instance, in paramagnetic phase of HoB$_{12}$ ($T_\textmd{N}$ $-$ 20~K) a short-range magnetic correlations between Ho-moments (one-dimensional spin chains) were detected along the [111] direction in neutron experiments \cite{89, 90}.

\section*{5. Conclusions}\label{Sec5}

Summarizing up, we have measured and analyzed the resistivity and transverse magnetoresistance of $R$$_x$Lu$_{1-x}$B$_{12}$ diluted systems with $x$ $\leq$ 0.03. Our results allow to exclude the contribution of additional channel of scattering caused by the formation of dynamic charge stripes. The data obtained do not contain any features at 150~K and there are no any deviations from Kohler's rule below this point. Instead the direction \textbf{H}$\|$[110] another one orientation \textbf{H}$\|$[111] appears to be distinguished demonstrating the trend to saturation behavior of transverse magnetoresistance (current geometry \textbf{I}$\|$[110]), that indicates the presence of only closed orbits. We argue, that charge-transport anisotropy is originated in $R$$_x$Lu$_{1-x}$B$_{12}$ due to the anisotropy of electron-phonon scattering on the one hand and the effects of Fermi surface topology on the other hand.

\section*{Acknowledgments}

The authors are grateful to Shared Facility Center of Prokhorov General Physics Institute of the RAS. We also express our thanks to K. Krasikov for useful discussions. One of us (N.Sh.) acknowledges Ulrich Burkhardt for the help in sample's attestation. This research was funded by the Russian Science Foundation Grant No. 22-12-00008 (https://rscf.ru/project/22-12-00008/).


\newpage
\begin{center}
    \textbf{Supplementary Materials to the article}\label{SecSM}
\end{center}
\setcounter{section}{-1}
\renewcommand{\thesection}{S.1}
\renewcommand{\thetable}{S\arabic{table}}
\renewcommand{\thefigure}{S\arabic{figure}}
\setcounter{figure}{0}
\setcounter{table}{0}

\section*{S.1. The characterization of the samples.}\label{SecS1}
\textbf{S.1.1.}\label{SecS1p1} Lu$_2$O$_3$ powder was preliminarily annealed at 800$^{\circ}$C for 2 hours to remove the crystallization water, after that the charge was prepared from oxide and boron according to the equation of a solid-state reaction of the borothermal reduction \cite{S1, S2}:

\begin{equation}\label{Eq.S1}
\textmd{Lu}_2\textmd{O}_3 + 27\textmd{B} \rightarrow 2\textmd{LuB}_{12} + \textmd{BO}\uparrow. \quad\quad\quad (\textmd{S}1)\nonumber
\end{equation}

\noindent
A small excess of boron (3 wt~$\%$) was introduced into the initial charge to compensate for possible boron losses due to its high vapor pressure at the synthesis temperature. The charge was mechanically mixed for several days, sifting through a sieve at least 5 times in order to break conglomerates based on oxide and boron and prepare, if possible, a homogeneous mixture. The prepared mixture was pressed into tablets with a diameter of 15~mm and a height of 10~mm, which were kept in a vacuum furnace for an hour at a temperature of 1650$^{\circ}$C. The temperature was raised slowly at a rate of $\sim$ 30~deg/min to ensure the removal of the evolved gases and the completion of the reaction, which is multistage, since the equation (\hyperref[Eq.S1]{S1}) reflects the overall reaction [initial (left side) and final (right side) stages], however, the reduction reaction itself goes through several intermediate stages, including the formation of borate and boride with a lower boron content and their subsequent interaction with the remaining boron to form dodecaboride \cite{S3, S4}:

\begin{equation}\label{Eq.S2}
\textmd{I}.\quad\textmd{Lu}_2\textmd{O}_3 + 5\textmd{B} \rightarrow \textmd{LuBO}_{3} + \textmd{LuB}_4\nonumber
\end{equation}
\begin{equation}\label{Eq.S3}
\textmd{II}.\quad\textmd{LuB}_4 + 8\textmd{B} \rightarrow \textmd{LuB}_{12}\nonumber
\end{equation}
\begin{equation}\label{Eq.S4}
\textmd{III}.\quad\textmd{LuBO}_3 + 14\textmd{B} \rightarrow \textmd{LuB}_{12} + 3\textmd{BO}\uparrow\nonumber
\end{equation}

\noindent
Since the solid-phase synthesis is determined by diffusion processes an additional homogenization of the synthesized powders was carried: the sintered tablets were broken, pressed again and kept in vacuum at 1750$^{\circ}$C for an hour. The annealed tablets were again broken; the powder was cold pressed into rods with a diameter of 8~mm and a length of 60~mm, which were sintered for one hour at 1750$^{\circ}$C in vacuum. Final homogenization occurs in the melt during crystal growth.

The synthesis, annealing of tablets and sintering of the rods took place in the same ZrB$_2$ crucibles. The melting temperature of ZrB$_2$ is $\sim$ 3000$^{\circ}$C; therefore, contamination with crucible material is excluded. Thus, the initial LuB$_{12}$ rods for crystal growth were identical in composition and purity. Volatile impurities present in boron are removed during synthesis and zone melting, and in the grown crystals the amount of impurities does not exceed 10$^{-2}$~wt~$\%$ (except for RE) according to the optical spectral analysis. Rare earth impurities are determined by the purity of the initial lutetium oxide, the total content of accompanying RE impurities did not exceed than 1.5$\times$10$^{-3}$~wt~$\%$.

Zone melting is carried out in a closed chamber under the pressure of high-purity argon (volume fraction of argon is not less than 99.993~$\%$). The process of the chamber preparing for melting is also identical for all crystals. Preliminarily the chamber is pumped out to 10$^{-3}$~mm~Hg (0.1333~Pa), then it is filled with argon up to 0.2~MPa and pumped out again, after which it is filled with argon to a predetermined pressure, and the melting process begin.
\vspace{1cm}

\textbf{S.1.2.}\label{SecS1p2} The samples quality and orientation were controlled by the X-ray diffraction and Laue backscattering patterns. X-ray phase analysis of the crushed LuB$_{12}$ single crystals (performed by using an X-ray diffractometer HZG-4A in Cu K$_{\alpha}$ radiation with a Ni filter) revealed reflections of the UB$_{12}$ structure type only (Fig.\hyperref[FigX1]{1d} in the main text). Any traces caused by the presence of borate and LuB$_4$ have never been detected (Fig.\hyperref[FigX1]{1d} in the main text). The presence of oxygen in the grown LuB$_{12}$ single crystals was checked by using pulsed reductive extraction with carbon in a flow of helium gas (the gas chromatography). The oxygen content did not exceed 0.04~mass~$\%$, $i.e.$ oxygen is chemically bound to the surface of the crushed single crystal.

Laue backscattering patterns (Figs.\hyperref[FigX1]{1f}, \hyperref[FigX1]{1h} in the main text and also Fig.S3) were taken in W or Co radiation. Fig.S3 shows the initial Tm$_{0.01}$Lu$_{0.99}$B$_{12}$ crystal with oriented plates which were cut out from it and their Laue backscattering patterns (Figs.S3a$-$S3c). There is no splitting of the point reflections on the X-ray Laue patterns, which confirms the absence of domains with a misorientation of more than tenths of a degree (procedure accuracy).

The real structure of the grown single crystals was studied by using scanning electron microscopy (SEM JSM-6490-LV, JEOL, Japan). Electron diffraction and high-resolution transmission microscopy analysis allowed us to determine the characteristics of local structural details of the single crystals. Electron Kikuchi patterns (Fig.\hyperref[FigX1]{1e} in the main text and Fig.S2c) and electron diffraction patterns obtained for LuB$_{12}$ single crystal (Fig.\hyperref[FigX1]{1c} in the main text and also Figs.S2a, S2b) showed that practically defect-free crystals were obtained. More details about the characterization of LuB$_{12}$ samples are presented in \cite{S1, S2, S5}.

\begin{figure}[b]
\begin{center}
\includegraphics[width = 9.5cm]{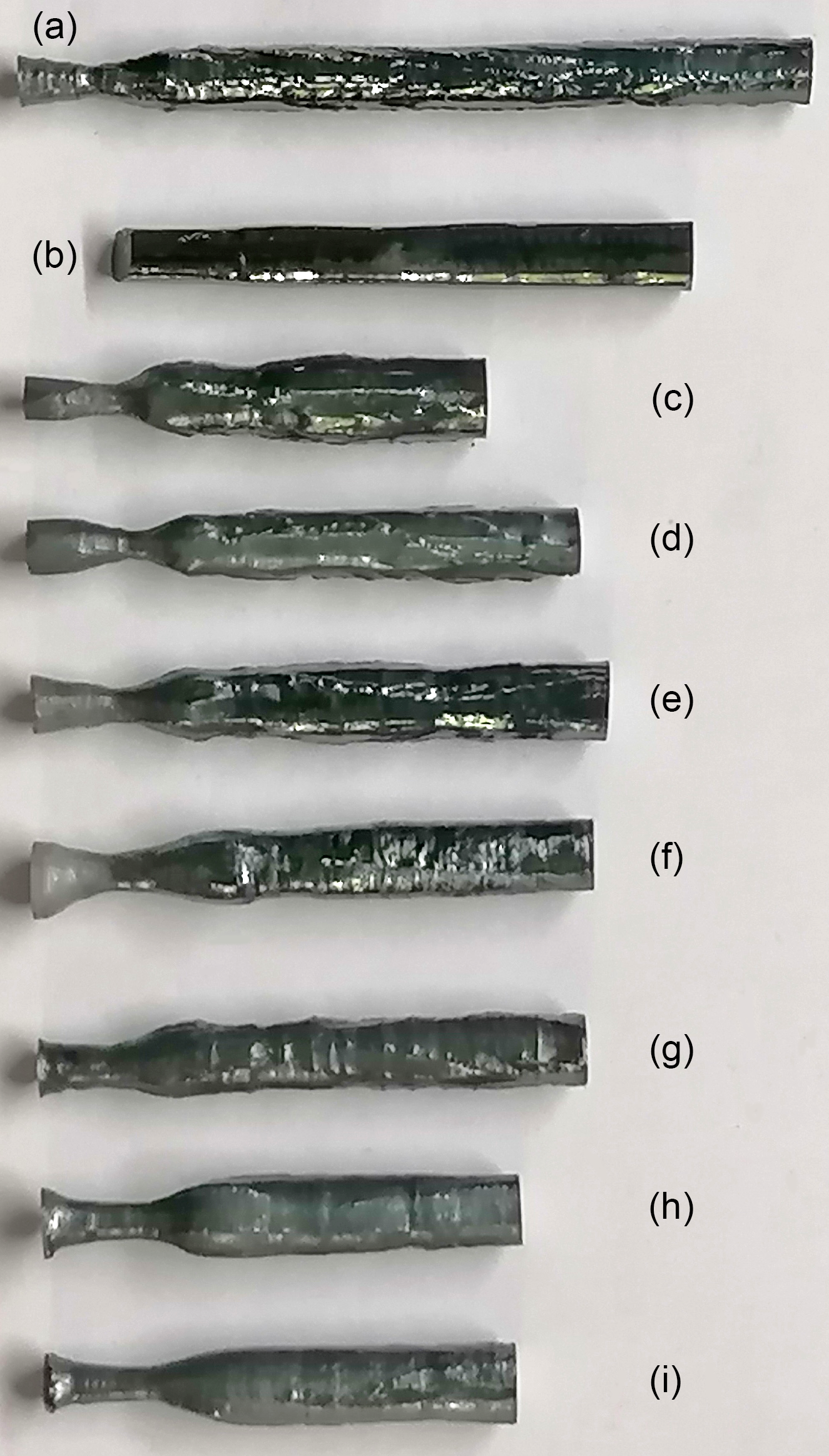}
  \caption{(Colour on-line).  The pictures of $R_x$Lu$_{1-x}$B$_{12}$ crystals as-grown including (a) LuB$_{12}$, (b) Dy$_{0.01}$Lu$_{0.99}$B$_{12}$, (c) Ho$_{0.01}$Lu$_{0.99}$B$_{12}$, (d) Er$_{0.01}$Lu$_{0.99}$B$_{12}$, (e) Tm$_{0.01}$Lu$_{0.99}$B$_{12}$, (f) Yb$_{0.01}$Lu$_{0.99}$B$_{12}$, (g) Tm$_{0.03}$Lu$_{0.97}$B$_{12}$, (h) Yb$_{0.02}$Lu$_{0.98}$B$_{12}$, (i) Yb$_{0.03}$Lu$_{0.97}$B$_{12}$ compositions.}\label{FigS1}
   \end{center}
\end{figure}

\newpage

   \begin{figure*}[!t]
\includegraphics[width = 16cm]{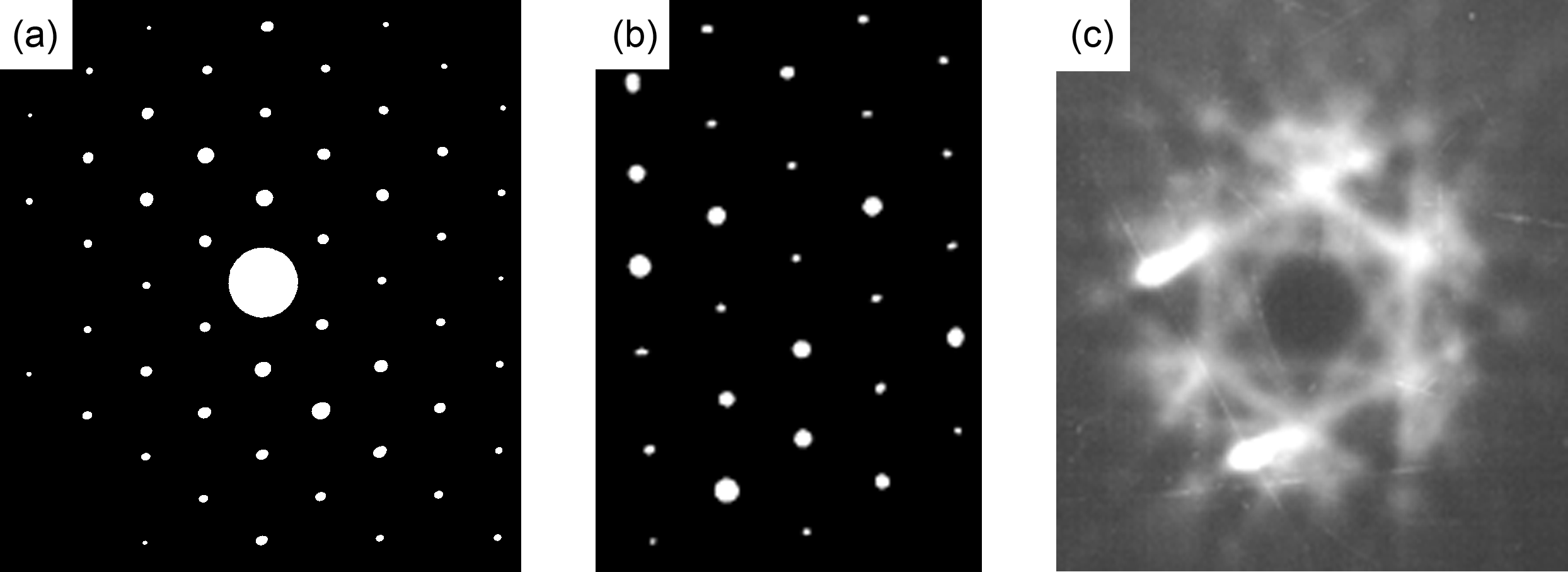}
   \parbox{18cm}{\caption{(Colour on-line). (a)-(b) Electron diffraction patterns (EDP) from local regions of LuB$_{12}$ single crystal along the other main crystallographic axis (a) [110], (b) [111]. EDP along [001] is presented for LuB$_{12}$ in Fig.\hyperref[FigX1]{1c} in the main text. (c) Symmetric electron Kikuchi pattern along the [111] direction in LuB$_{12}$ single crystal. Symmetric electron Kikuchi pattern along the [100] direction is presented for LuB$_{12}$ in Fig.\hyperref[FigX1]{1e} in the main text.}}\label{FigS2}
\end{figure*}

\begin{figure}[!b]
\begin{center}
\includegraphics[width = 14.2cm]{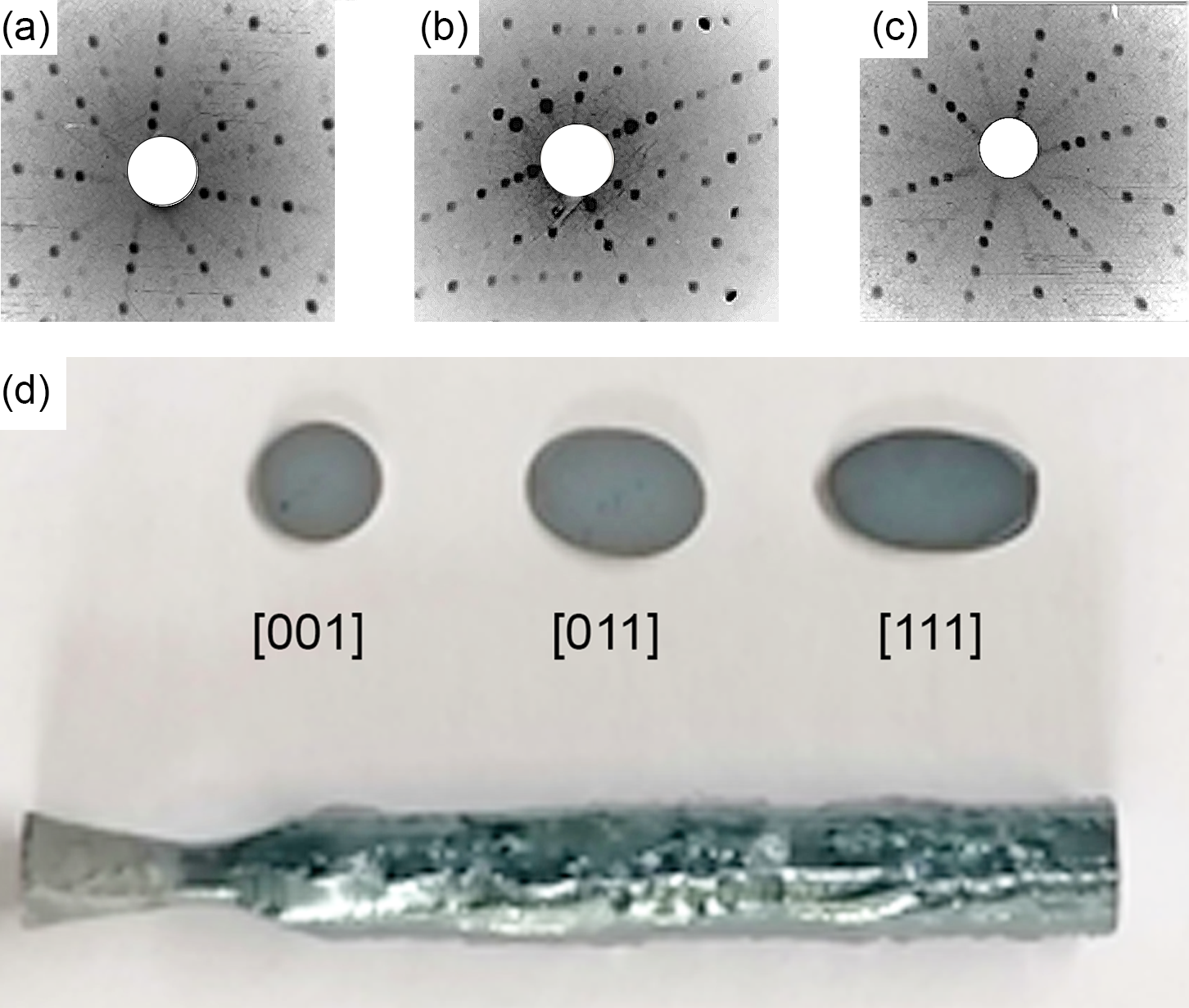}
  \caption{(Colour on-line).  (a)-(c) Typical X-ray Laue backscattering patterns of the oriented plates prepared from the same single crystal of Tm$_{0.01}$Lu$_{0.99}$B$_{12}$ including (a) [001], (b) [011], and (c) [111]. (d) The picture of Tm$_{0.01}$Lu$_{0.99}$B$_{12}$ single crystal as-grown with corresponding oriented plates, which were cut out from it.}\label{FigS3}
   \end{center}
\end{figure}

\newpage

\section*{S2. Details about the transport properties of $R_x$Lu$_{1-x}$B$_{12}$ and the analysis of zero-field resistivity.}\label{SecS2}

Here we present a complete set of $\Delta\rho/\rho$ = $f$($H$) experimental data of the systems under investigation (Figs.S6$-$S8) and also the comparison with results obtained by our group previously for Ho$_{0.01}$Lu$_{0.99}$B$_{12}$ and host Lu$^{\textmd{nat}}$B$_{12}$, La$^{\textmd{nat}}$B$_6$ objects (Figs.S4$-$S5, S9$-$S11). For analysis of zero-field resistivity in ZrB$_{12}$ and UB$_{12}$ (Table S1) corresponding data were digitized from the literature.

\vspace{1cm}

\begin{figure}[htpb]
\begin{minipage}[h]{0.37\linewidth}
 \center{\includegraphics[width=1\linewidth]{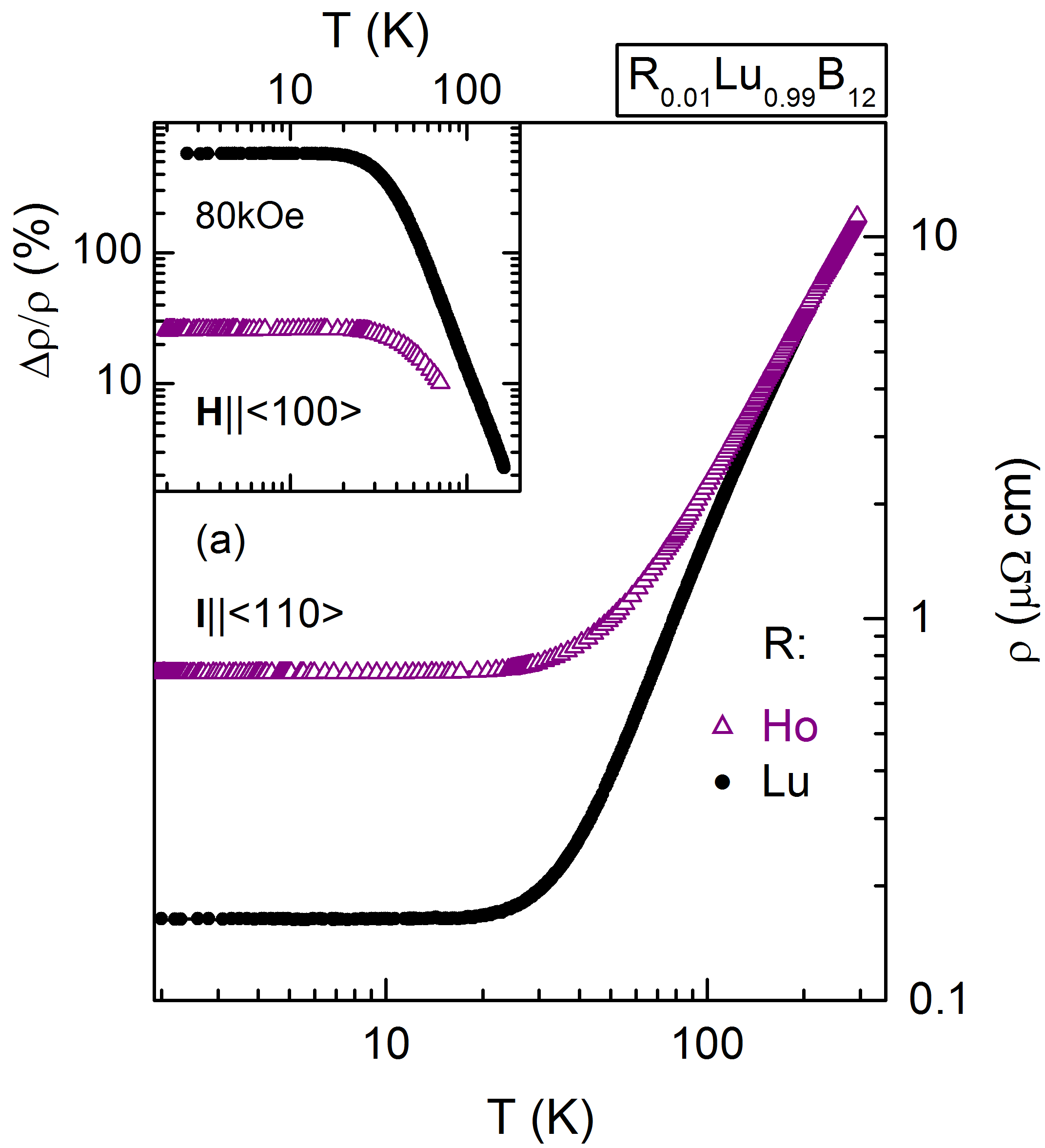}}
\end{minipage}
\begin{minipage}[h]{0.37\linewidth}
 \center{\includegraphics[width=1\linewidth]{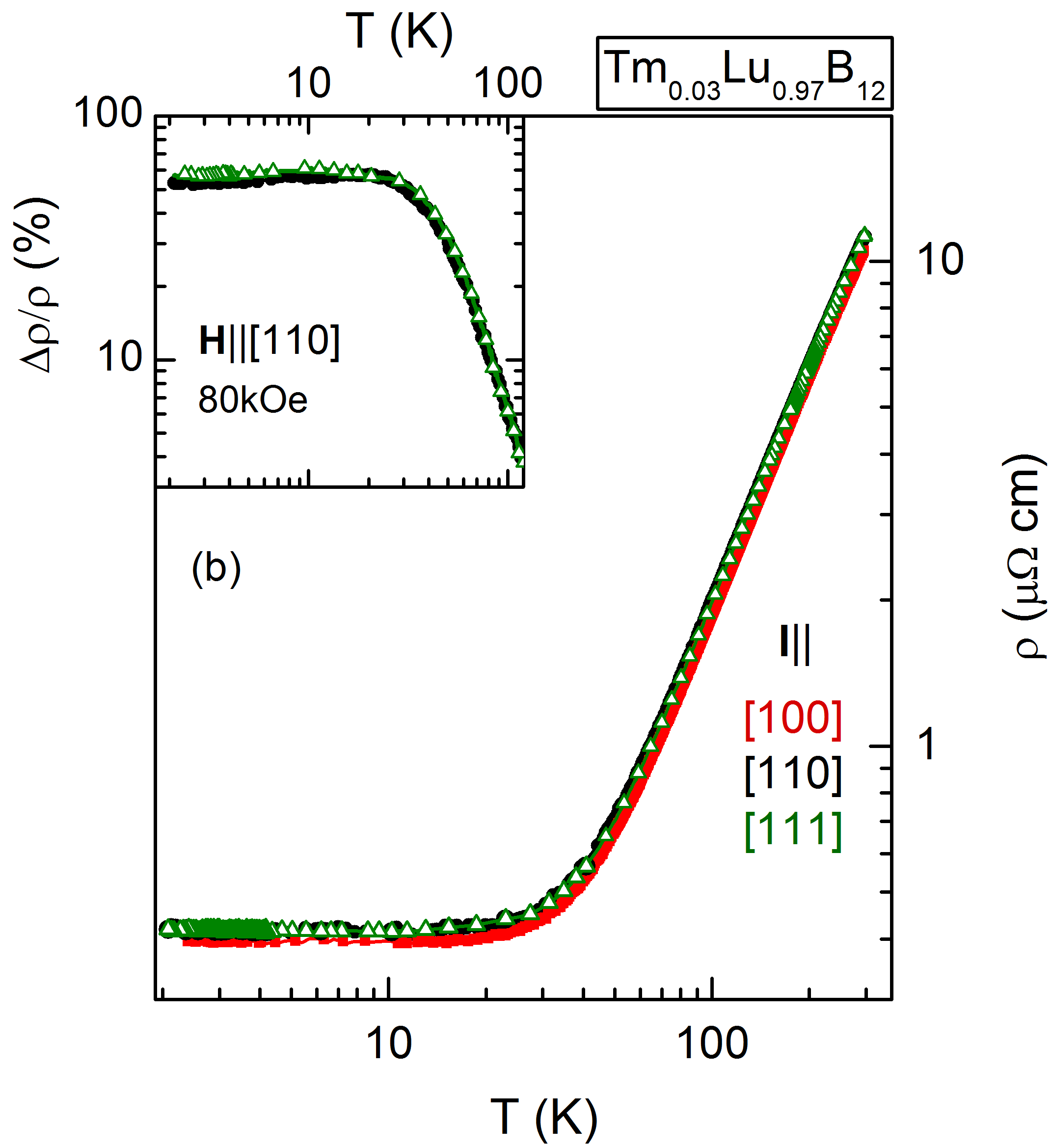}}
\end{minipage}
  \caption{(Colour on-line). Temperature evolution of zero-field electrical resistivity $\rho$($T$) of (a) Ho$_{0.01}$Lu$_{0.99}$B$_{12}$ and LuB$_{12}$ objects (experimental geometry \textbf{I}$\|$$<$110$>$, \textbf{H}$\|$$<$100$>$). Panel (b) displays $\rho$($T$) data measured for Tm$_{0.03}$Lu$_{0.97}$B$_{12}$ crystals prepared with three different current orientations \textbf{I}$\|$[100], \textbf{I}$\|$[110] and \textbf{I}$\|$[111]. Insets in each panel show $T$-dependence of transverse magnetoresistance $\Delta\rho/\rho$ = $f$($T$) at constant magnetic field $H$ = 80~kOe. The data on panel (a) were taken from \cite{S6}. Additional information about Ho$_x$Lu$_{1-x}$B$_{12}$ single crystals quality is presented in SM of \cite{S6} and also in \cite{S1,S2}.}\label{FigS4}
\end{figure}

\begin{figure}[!b]
\begin{center}
\includegraphics[width = 5cm]{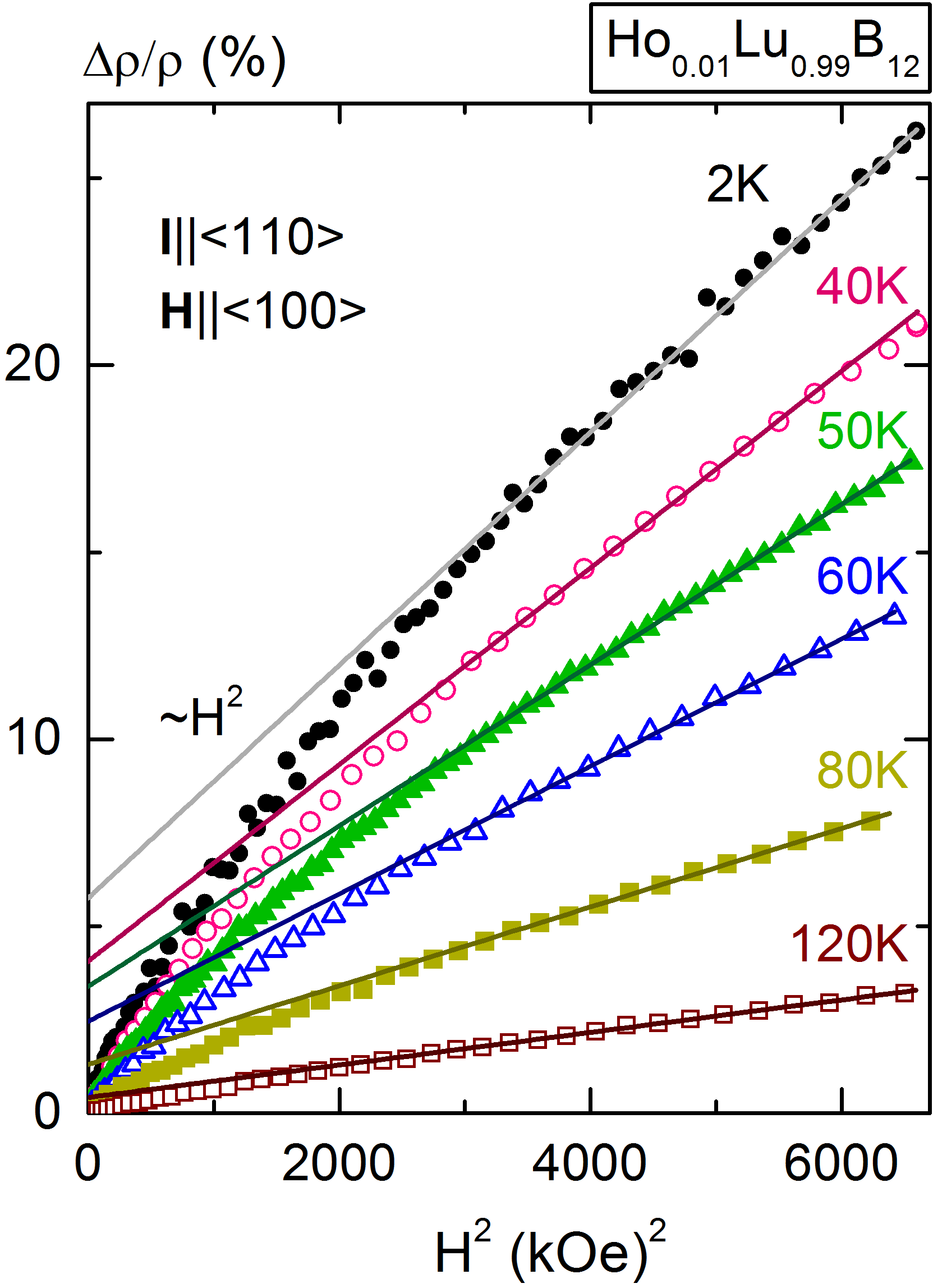}
  \caption{(Colour on-line). Field scans of magnetoresistance presented in coordinates $\Delta\rho/\rho$ = $f$($H^2$, $T_0$) for Ho$_{0.01}$Lu$_{0.99}$B$_{12}$ compound. Solid lines represent quadratic asymptotic $\sim$ $H^2$. The data were taken from \cite{S6}. The geometry of the experiment was \textbf{I}$\|$$<$110$>$ and \textbf{H}$\|$$<$100$>$ \cite{S6}.}\label{FigS5}
   \end{center}
\end{figure}

\newpage
\LTcapwidth=16cm
\begin{longtable*}[!t]{ccccccccc}
\caption{Analysis summary of zero-field resistivity data digitized for ZrB$_{12}$ and UB$_{12}$ from \cite{S7} and \cite{S8}, respectively (see main text for details). Here $\rho_0$ designates residual resistivity; $\lambda_{\textmd{tr}}$$\omega_\textmd{D}$/$\omega_\textmd{P}$$^2$ and ($KN$/$m$)$_i$ are amplitude factors in Eq.(\hyperref[Eq.2]{2}) and Eq.(\hyperref[Eq.3]{3}) in the main text, respectively; $\Theta_{\textmd{D}}$ and $\Theta_{\textmd{Ei}}$ are Debye and Einstein temperatures; A$_{\textmd{U}}$ and $T_0$ are amplitude factor and energy of substantial phonons in Eq.(\hyperref[Eq.4]{4}) in the main text; $T_{00}$ is crossover point from $N$- to $U$-processes (see the main text for details).\label{TabS1}}\\
\hhline{=========} \\
  \quad\quad\quad \multirow{3}{*}{$R$B$_{12}$}  \quad\quad & \multirow{3}{*}{models}  \quad\quad&    $\rho_0$  \quad\quad& ($KN$/$m$)$_1$ \quad\quad &  $\Theta_{\textmd{E1}}$  \quad\quad& ($KN$/$m$)$_2$  \quad\quad& $\Theta_{\textmd{E2}}$ \quad\quad& $\lambda_{\textmd{tr}}$$\omega_\textmd{D}$/$\omega_\textmd{P}$$^2$ \quad\quad& $\Theta_{\textmd{D}}$  \quad\quad\quad\\
			\\
			\quad\quad\quad	\quad\quad &  \quad\quad  &    ($\mu$$\Omega$$\cdot$cm) \quad\quad & (m$\Omega$$\cdot$cm$\cdot$K) \quad\quad &  (K) \quad\quad& (m$\Omega$$\cdot$cm$\cdot$K) \quad\quad& (K) \quad\quad& ($\mu$$\Omega$$\cdot$cm) \quad\quad& (K) \quad\quad\quad\\
			\\
\hhline{---------} \\
\\
\quad\quad\quad ZrB$_{12}$ \quad\quad & model I  \quad\quad &   1.8 \quad\quad  & 1.61  \quad\quad &  195 \quad\quad & $-$ \quad\quad & $-$ \quad\quad & 0.233 \quad\quad & 1160 \quad\quad\quad \\
& single Einstein component &    &   &   &  &  &  &     \\
\\
\\
\quad\quad\quad			UB$_{12}$	\quad\quad & model I   \quad\quad &    0.726  \quad\quad & 2.1  \quad\quad &  237 \quad\quad & $-$ \quad\quad & $-$ \quad\quad & 1.62 \quad\quad & 1160  \quad\quad\quad \\
& single Einstein component &    &   &   &  &  &  &   \\
   \\
\quad\quad\quad			UB$_{12}$	\quad\quad & model II  \quad\quad &  0.726  \quad\quad &  0.72  \quad\quad &    178 \quad\quad & 1.65 \quad\quad & 347 \quad\quad & 1.7 \quad\quad & 1160  \quad\quad\quad \\
& two Einstein components &    &   &   &  &  &  &     \\						
      \\
		\hhline{=========}
\end{longtable*}

\begin{figure}[!b]
\begin{center}
\includegraphics[width = 14cm]{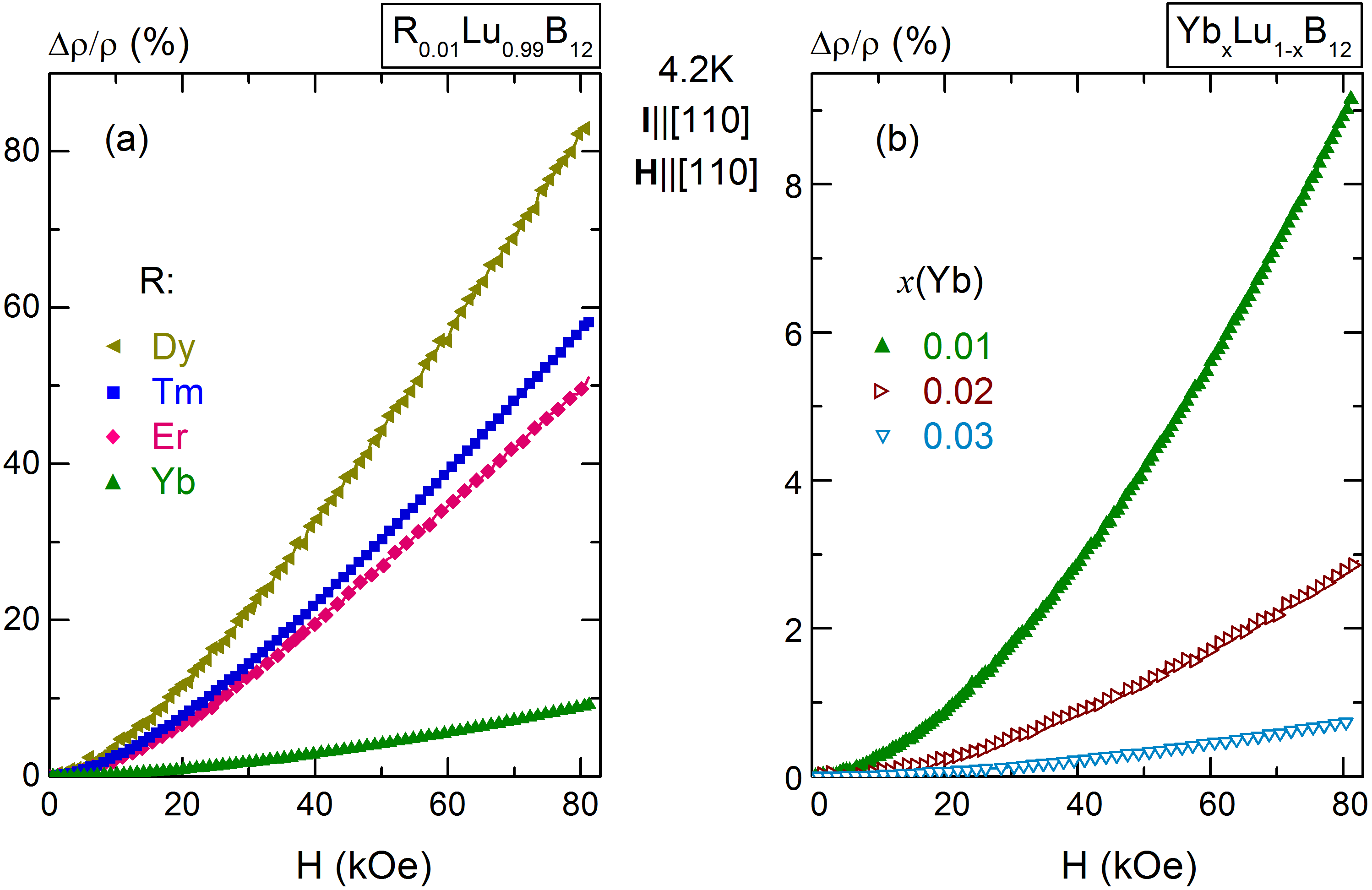}
  \caption{(Colour on-line). Field scans of transverse magnetoresistance $\Delta\rho/\rho$ = $f$($H$, $T_0$) for (a) $R_{0.01}$Lu$_{0.99}$B$_{12}$ ($R$-Dy, Er, Tm, and Yb) diluted systems, and separately for (b) Yb$_x$Lu$_{1-x}$B$_{12}$ family at $T$ = 4.2~K. Experimental geometry was \textbf{I}$\|$[110], and \textbf{H}$\|$[110].}\label{FigS6}
   \end{center}
\end{figure}

\newpage

     \begin{figure*}[!t]
\includegraphics[width = 18.5cm]{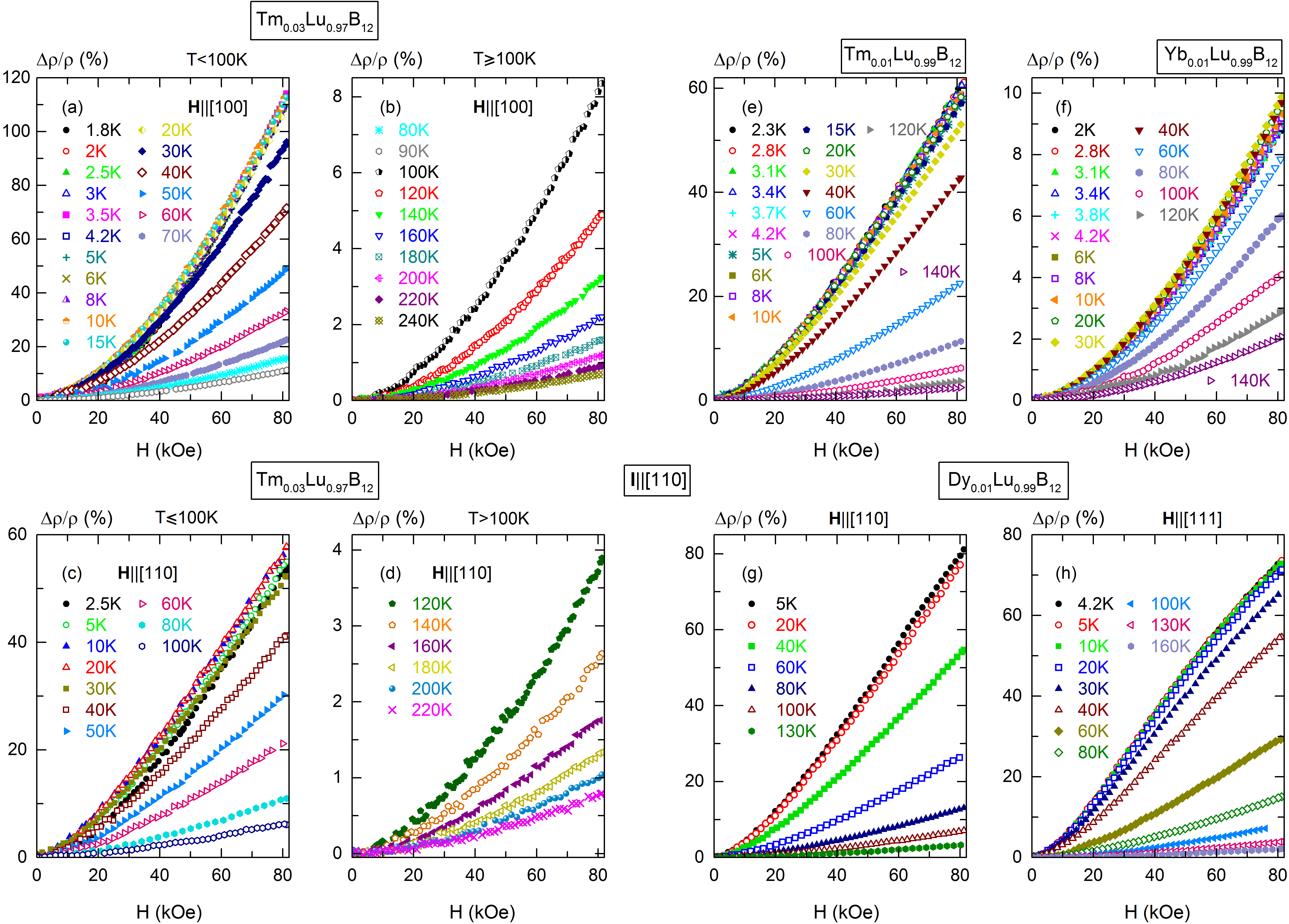}
  \hspace{-2cm} \parbox{18cm}{\caption{(Colour on-line). Field dependencies of transverse magnetoresistance $\Delta\rho/\rho$ = $f$($H$, $T_0$) for (a)-(d) Tm$_{0.03}$Lu$_{0.97}$B$_{12}$, (e) Tm$_{0.01}$Lu$_{0.99}$B$_{12}$, (f) Yb$_{0.01}$Lu$_{0.99}$B$_{12}$ and (g)-(h) Dy$_{0.01}$Lu$_{0.99}$B$_{12}$. Different field geometries (\textbf{H}$\|$[100], \textbf{H}$\|$[110], \textbf{H}$\|$[111]) with the same  current orientation (\textbf{I}$\|$[110]) are presented for Tm$_{0.03}$Lu$_{0.97}$B$_{12}$ and Dy$_{0.01}$Lu$_{0.99}$B$_{12}$ compounds.}}\label{FigS7}
\end{figure*}

$\phantom{x}$

$\phantom{x}$

$\phantom{x}$

$\phantom{x}$

$\phantom{x}$

$\phantom{x}$

$\phantom{x}$

$\phantom{x}$

$\phantom{x}$

$\phantom{x}$

$\phantom{x}$

$\phantom{x}$

$\phantom{x}$

$\phantom{x}$

$\phantom{x}$

$\phantom{x}$

\newpage
      \begin{figure*}[!t]
\includegraphics[width = 14cm]{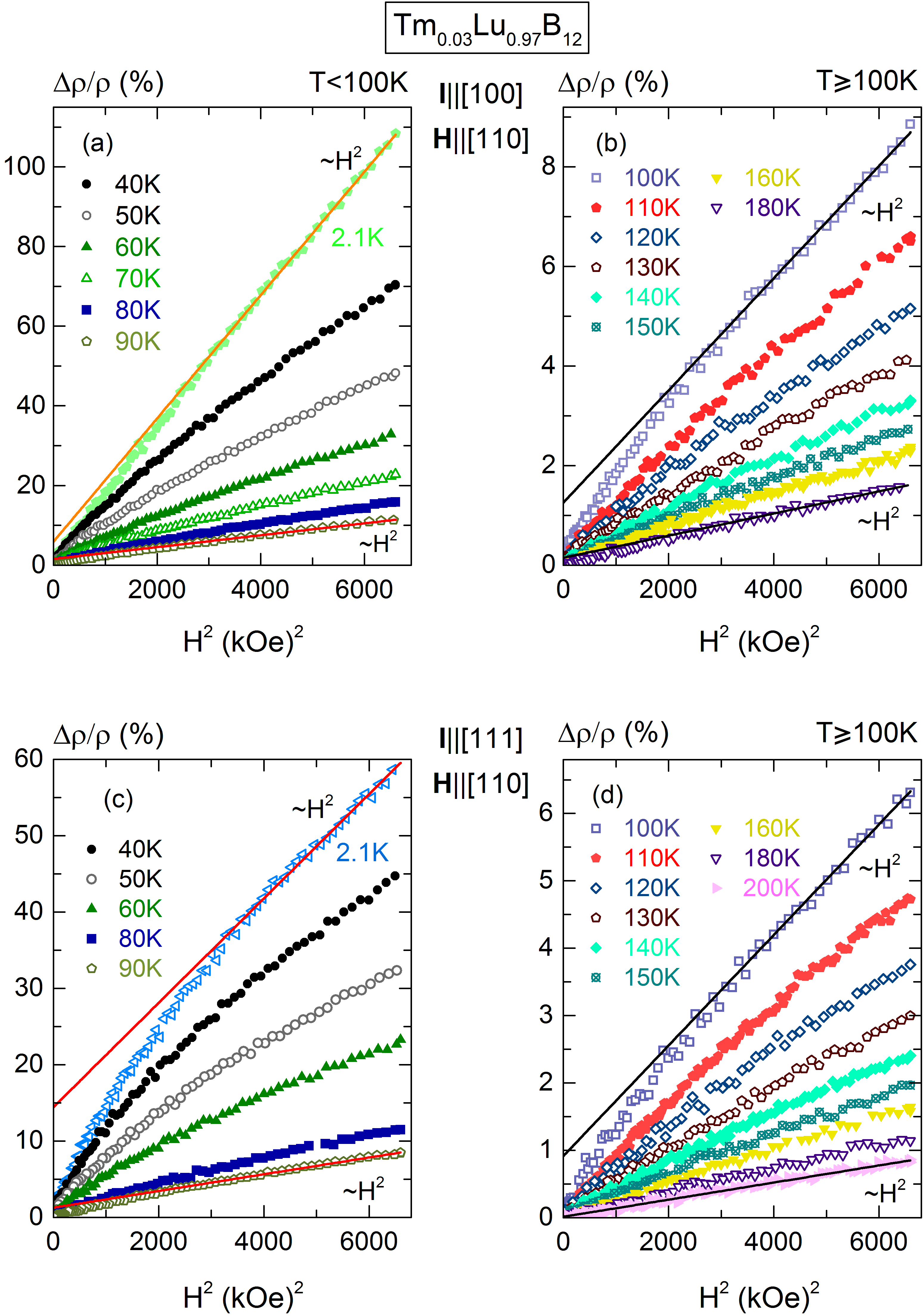}
   \parbox{18cm}{\caption{(Colour on-line). Field scans of magnetoresistance in coordinates $\Delta\rho/\rho$ = $f$($H^2$, $T_0$) measured for Tm$_{0.03}$Lu$_{0.97}$B$_{12}$ at various current geometries (a)-(b) \textbf{I}$\|$[100] and (c)-(d) \textbf{I}$\|$[111]  with the same field orientation
   \textbf{H}$\|$[110].  Solid lines represent quadratic asymptotic $\sim$ $H^2$.}}\label{FigS8}
\end{figure*}

\newpage

\LTcapwidth=16cm
\begin{longtable*}[!t]{ccccccccccccc}
\caption{Analysis summary for Lu$^{\textmd{nat}}$B$_{12}$ from Fig.S9 by Eqs.(\hyperref[Eq.1]{1})-(\hyperref[Eq.3]{3}) ($N$-processes only) and Eq.(\hyperref[Eq.1]{1}), Eqs.(\hyperref[Eq.3]{3})-(\hyperref[Eq.4]{4}) (with $U$-processes): $\rho_0$ is residual resistivity; $\lambda_{\textmd{tr}}$$\omega_\textmd{D}$/$\omega_\textmd{P}$$^2$ and ($KN$/$m$)$_i$ are amplitude factors in Eq.(\hyperref[Eq.2]{2}) and Eq.(\hyperref[Eq.3]{3}) in the main text, respectively; $\Theta_{\textmd{D}}$ and $\Theta_{\textmd{Ei}}$ are Debye and Einstein temperatures; A$_{\textmd{U}}$ and $T_0$ are amplitude factor and energy of substantial phonons in Eq.(\hyperref[Eq.4]{4}) in the main text; $T_{00}$ is crossover point from $N$- to $U$-processes (see the main text for details).\label{TabS2}}\\
\hhline{=============} \\

		\quad	\multirow{3}{*}{models}  \quad&    $\rho_0$  \quad& ($KN$/$m$)$_1$  \quad&  $\Theta_{\textmd{E1}}$  \quad& ($KN$/$m$)$_2$  \quad& $\Theta_{\textmd{E2}}$ \quad& $\lambda_{\textmd{tr}}$$\omega_\textmd{D}$/$\omega_\textmd{P}$$^2$ \quad& $\Theta_{\textmd{D}}$ \quad& A$_{\textmd{U}}$ \quad& T$_{0}$ \quad& T$_{00}$\quad\\
\\
				\quad    \quad&    ($\mu$$\Omega$$\cdot$cm)  \quad& (m$\Omega$$\cdot$cm$\cdot$K)  \quad&  (K) \quad& (m$\Omega$$\cdot$cm$\cdot$K) \quad& (K) \quad& ($\mu$$\Omega$$\cdot$cm) \quad& (K) \quad& (n$\Omega$$\cdot$cm/K)  \quad& (K) \quad& (K) \quad\\
			\\
\hhline{-------------} \\
\\
		\quad model I (Fig.S9a) \quad &   0.164  \quad & 0.552  \quad &  200 \quad & $-$ \quad & $-$ \quad & 0.667 \quad & 1160 \quad & $-$ \quad & $-$ \quad & $-$ \quad\\
\quad single Einstein component \quad &   \quad &   \quad&  \quad &  \quad&  \quad&  \quad&  \quad& \quad & \quad & \quad\\
\\
			\quad	model I (Fig.S9b)  \quad&    0.164  \quad& 0.157  \quad&  150 \quad& 2.52\quad & 365 \quad& 0.455 \quad& 1160 \quad& $-$ \quad& $-$ \quad& $-$\quad\\
two Einstein components \quad&    \quad&   \quad&  \quad &  \quad&  \quad& \quad &  \quad&  \quad&  \quad& \quad\\
   \\
			\quad	model II (Fig.S9c) \quad&   0.164  \quad&  0.188  \quad&    168.5 \quad& $-$ \quad& $-$ \quad& $-$ \quad& $-$ \quad& 57 \quad& 172\quad & 69\quad \\
single Einstein component \quad&    \quad&  \quad & \quad  &  \quad&  \quad&  \quad&  \quad&  \quad&  \quad& \quad \\			
	\\
		\hhline{=============}
\end{longtable*}

         \begin{figure*}[!b]
\includegraphics[width = 18cm]{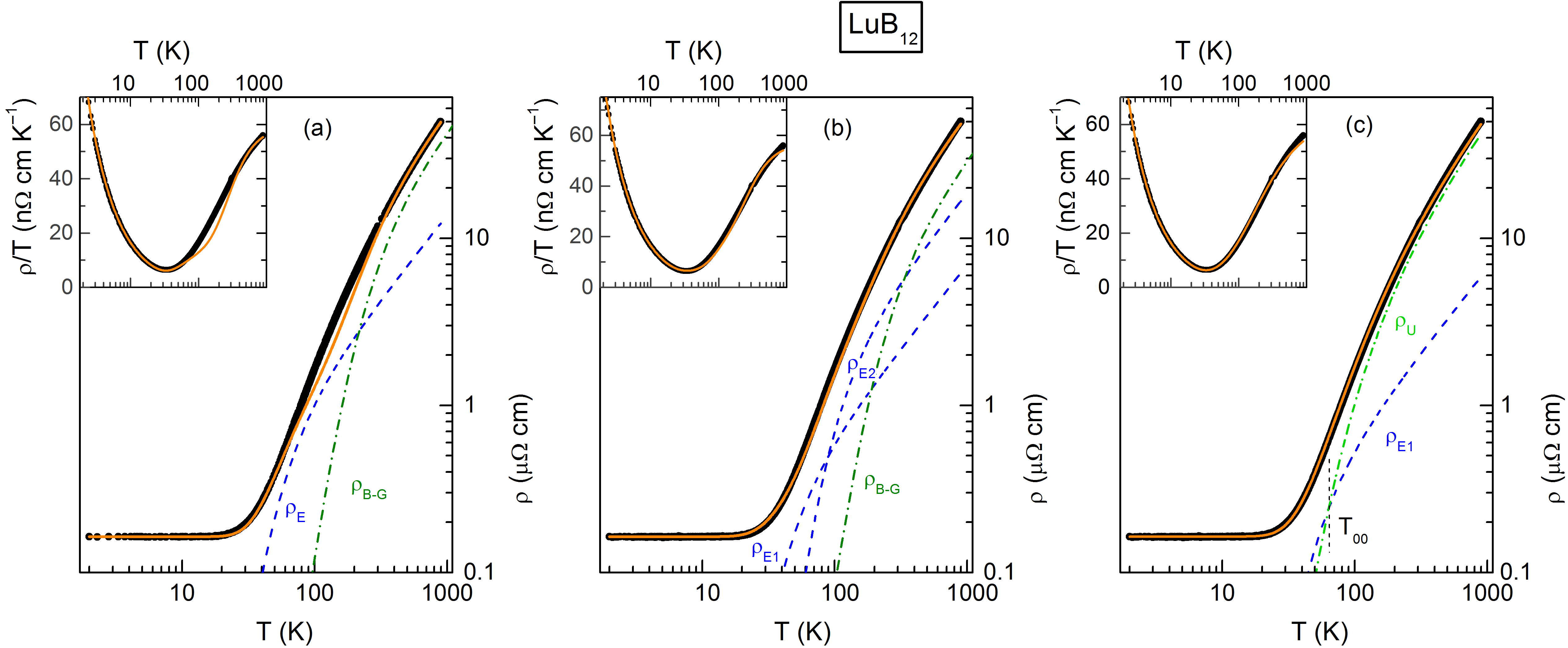}
 \hspace{-1cm}  \parbox{18cm}{\caption{(Colour on-line). The analysis of zero field electrical resistivity $\rho$($T$) of Lu$^{\textmd{nat}}$B$_{12}$ in the framework of model \cite{S9} by using (a) single Einstein ($\Theta_{\textmd{E}}$ $\approx$ 200~K) and (b) two Einstein components ($\Theta_{\textmd{E1}}$ $\approx$ 150~K, $\Theta_{\textmd{E2}}$ $\approx$ 365~K) with Debye temperature $\Theta_{\textmd{D}}$ $\approx$ 1160~K (model I), and (c) model II with $T_0$ = $\Theta_{\textmd{E}}$ $\approx$  168.5~K ($T_{00}$ $\approx$ 69~K). Original resistivity data were taken from \cite{S10}. Other parameters of fitting are presented in Table S2. The inset shows the same approximation in coordinates $\rho$/$T$ vs $T$ (see the main text for details).}}\label{FigS9}
\end{figure*}

\newpage

\LTcapwidth=16cm
\begin{longtable*}[!t]{ccccccccccccc}
\caption{Analysis summary for La$^{\textmd{nat}}$B$_{6}$ from Fig.S10 by Eqs.(\hyperref[Eq.1]{1})-(\hyperref[Eq.3]{3}) ($N$-processes only) and Eq.(\hyperref[Eq.1]{1}), Eqs.(\hyperref[Eq.3]{3})-(\hyperref[Eq.4]{4}) (with $U$-processes): $\rho_0$ is residual resistivity; $\lambda_{\textmd{tr}}$$\omega_\textmd{D}$/$\omega_\textmd{P}$$^2$ and ($KN$/$m$)$_i$ are amplitude factors in Eq.(\hyperref[Eq.2]{2}) and Eq.(\hyperref[Eq.3]{3}) in the main text, respectively; $\Theta_{\textmd{D}}$ and $\Theta_{\textmd{Ei}}$ are Debye and Einstein temperatures; A$_{\textmd{U}}$ and $T_0$ are amplitude factor and energy of substantial phonons in Eq.(\hyperref[Eq.4]{4}) in the main text; $T_{00}$ is crossover point from $N$- to $U$-processes (see the main text for details).\label{TabS3}}\\
\hhline{=============} \\

		\quad	\multirow{3}{*}{models}  \quad&    $\rho_0$  \quad& ($KN$/$m$)$_1$  \quad&  $\Theta_{\textmd{E1}}$  \quad& ($KN$/$m$)$_2$  \quad& $\Theta_{\textmd{E2}}$ \quad& $\lambda_{\textmd{tr}}$$\omega_\textmd{D}$/$\omega_\textmd{P}$$^2$ \quad& $\Theta_{\textmd{D}}$ \quad& A$_{\textmd{U}}$ \quad& T$_{0}$ \quad& T$_{00}$\quad\\
\\
				\quad    \quad&    ($\mu$$\Omega$$\cdot$cm)  \quad& (m$\Omega$$\cdot$cm$\cdot$K)  \quad&  (K) \quad& (m$\Omega$$\cdot$cm$\cdot$K) \quad& (K) \quad& ($\mu$$\Omega$$\cdot$cm) \quad& (K) \quad& (n$\Omega$$\cdot$cm/K)  \quad& (K) \quad& (K) \quad\\
			\\
\hhline{-------------} \\
\\
		\quad	model I (Fig.S10a)\quad &   0.017  \quad& 0.543  \quad&  175 \quad& $-$ \quad& $-$ \quad& 0.215 \quad& 1160 \quad& $-$ \quad& $-$ \quad& $-$\quad\\
\quad single Einstein component \quad &   \quad &   \quad&   \quad&  \quad&  \quad& \quad &  \quad&  \quad&  \quad& \quad\\
\\
		\quad		model I (Fig.S10b)  \quad&    0.017  \quad& 0.347  \quad&  155 \quad& 0.55 \quad& 340 \quad& 0.185 \quad& 1160 \quad& $-$ \quad& $-$ \quad& $-$\quad\\
\quad two Einstein components \quad &    \quad &  \quad  &   \quad &  \quad & \quad &  \quad&  \quad& \quad &  \quad& \quad\\
   \\
			\quad	model II (Fig.S10c) \quad&  0.017  \quad&  0.257  \quad&    155 \quad& $-$ \quad& $-$ \quad& $-$ \quad& $-$ \quad& 23 \quad& 155 \quad& 189 \quad\\
\quad single Einstein component \quad &    \quad &   \quad &  \quad &  \quad&  \quad&  \quad& \quad & \quad &  \quad & \quad\\					
      	\\
		\hhline{=============}
\end{longtable*}

 \vspace{2cm}

         \begin{figure*}[!b]
\includegraphics[width = 18cm]{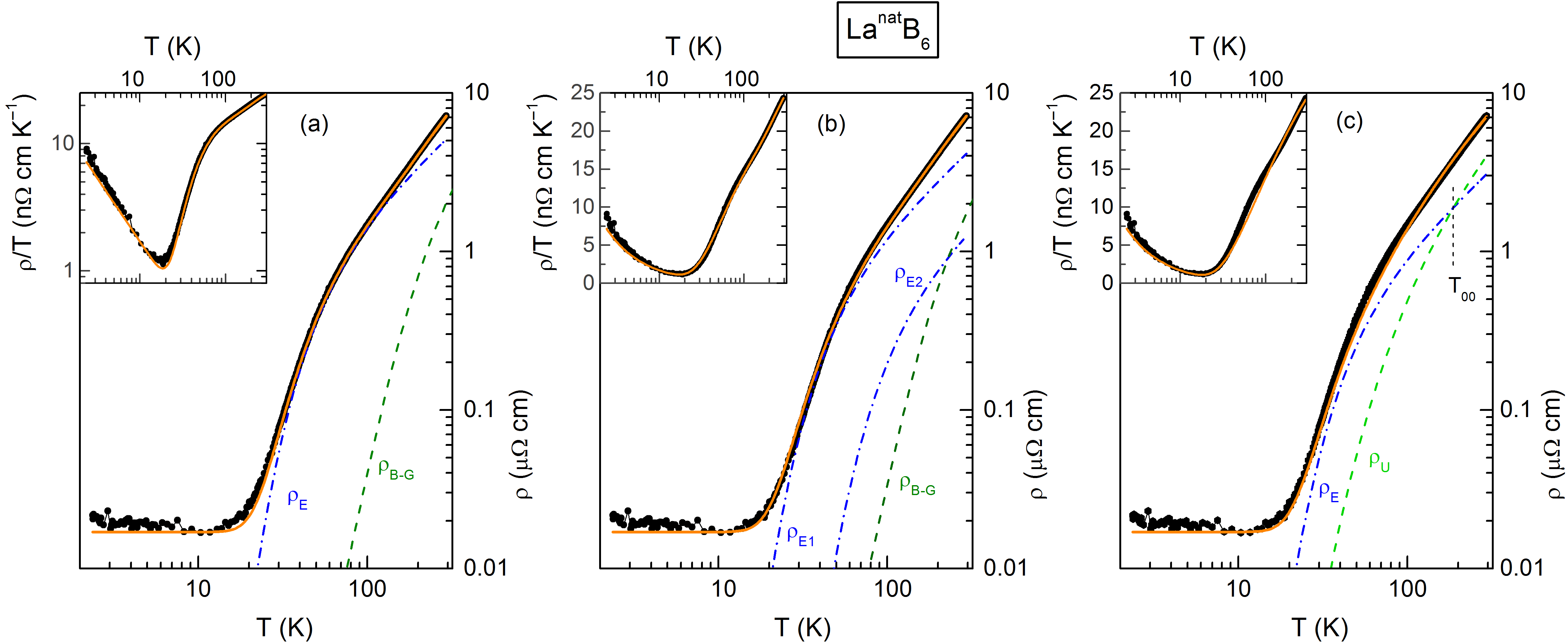}
 \hspace{-1cm}  \parbox{18cm}{\caption{(Colour on-line). The analysis of zero field electrical resistivity $\rho$($T$) of La$^{\textmd{nat}}$B$_6$ in the framework of model \cite{S9} by using (a) single Einstein ($\Theta_{\textmd{E}}$ $\approx$ 175~K) and (b) two Einstein components ($\Theta_{\textmd{E1}}$ $\approx$  155~K, $\Theta_{\textmd{E2}}$ $\approx$ 340~K) with Debye temperature $\Theta_{\textmd{D}}$ $\approx$ 1160~K (model I), and (c) model II with $T_0$ = $\Theta_{\textmd{E}}$ $\approx$ 155~K ($T_{00}$ $\approx$ 189~K). Original resistivity data were taken from \cite{S11}. Other parameters of fitting are presented in Table S3. The inset shows the same approximation in coordinates $\rho$/$T$ vs $T$ (see the main text for details).}}\label{FigS10}
\end{figure*}

\newpage

\begin{figure}[t]
\begin{center}
\includegraphics[width = 10cm]{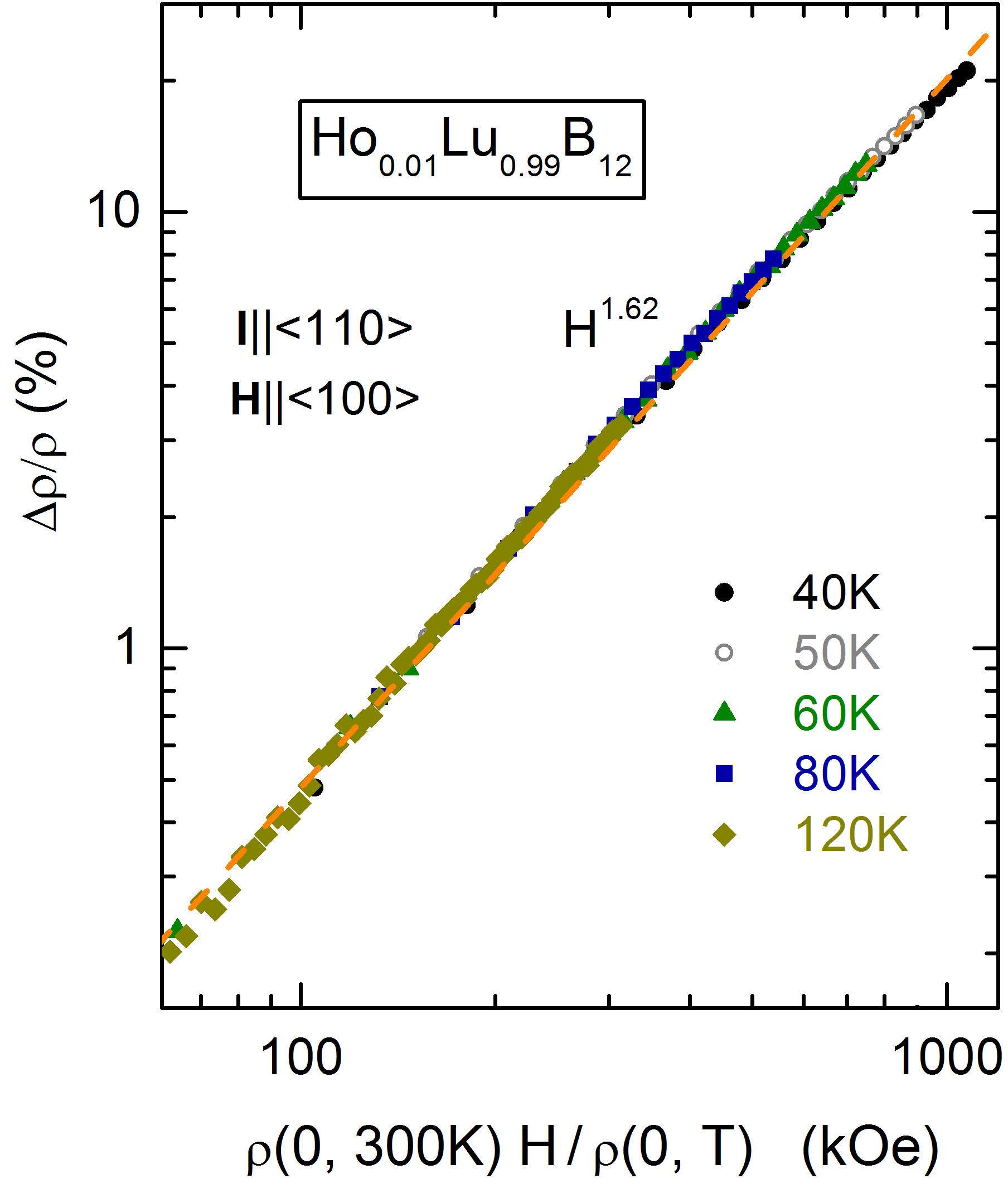}
  \caption{(Colour on-line). Kohler's plot of  magnetoresistance $\Delta\rho/\rho$ = $f$[$\rho$(0, 300~K)$H$/$\rho$(0, $T$)] for Ho$_{0.01}$Lu$_{0.99}$B$_{12}$ composition at various fixed temperatures from the range 40 $-$ 120~K. Experimental geometry was \textbf{I}$\|$$<$110$>$ and \textbf{H}$\|$$<$100$>$. Dashed lines display the approximation by power law $\Delta\rho/\rho$ $\sim$ $H^a$ with the index $a$ $\approx$ 1.62. The data for Ho$_{0.01}$Lu$_{0.99}$B$_{12}$ were taken from \cite{S6}.}\label{FigS11}
   \end{center}
\end{figure}


\begin{thebibliography}{999}
\bibitem{10} The transition to superconducting state has been reported for LaB$_6$ in literature at $T_\textmd{c}$ $\approx$ 0.1~K. Later alternative transport measurements performed on LaB$_6$ single crystals (prepared by using floating zone melting in IPM NASU, Kyiv) showed that normal state stays in this material at the temperatures down to 10~mK \cite{2}. Among the other reasons this feature may be caused by the effect when the superconductivity is observed in nonstoichiometric samples. Similar effect was discussed in literature for some borides such as ZrB$_2$, Mo$_2$B, W$_2$B,  etc.
\end{thebibliography}
\end{document}